\documentclass[a4paper, 10 pt, conference]{IEEEtran}  

\IEEEoverridecommandlockouts   
\usepackage{lipsum}
\usepackage{graphicx}
\usepackage{caption}
\usepackage{subcaption}
\usepackage{float}
\usepackage{multicol,multirow}
\usepackage{amsmath,amssymb,amsfonts}
\usepackage{amsthm}
\usepackage{rotating}
\usepackage[numbers]{natbib}
\usepackage{ifpdf}
\usepackage[T1]{fontenc}
\usepackage{sourcesanspro}%
\usepackage{textcomp}%
\usepackage{xcolor}
\usepackage{hyperref}
\hypersetup{
    colorlinks,
    linkcolor={red!50!black},
    citecolor={blue!50!black},
    urlcolor={blue!80!black}
}
\usepackage{pgfplots}
\pgfplotsset{compat=1.18}
\usepackage[none]{hyphenat}
\usepackage{parskip}
\usepackage{booktabs} 
\usepackage{xurl}
\usepackage{cleveref}
\usepackage{domitian}

\usepackage{titlesec}
\usepackage{blindtext}

\titlespacing*{\section}{0pt}{1.5\baselineskip}{1\baselineskip}
\titlespacing*{\subsection}{0pt}{1.25\baselineskip}{0.75\baselineskip}
\titlespacing*{\subsubsection}{0pt}{1.25\baselineskip}{0.75\baselineskip}
\def\BibTeX{{\rm B\kern-.05em{\sc i\kern-.025em b}\kern-.08em
    T\kern-.1667em\lower.7ex\hbox{E}\kern-.125emX}}

\title{\LARGE \bf Take the Train: Africa at the Crossroad of Modern AI }

\author{\IEEEauthorblockN{Cédric Manouan\IEEEauthorrefmark{1},
Miquilina Anagbah\IEEEauthorrefmark{4},
N'guessan Yves-Roland Douha\IEEEauthorrefmark{2}$^{,}$\IEEEauthorrefmark{3}, and
João Barros\IEEEauthorrefmark{2}$^{,}$\IEEEauthorrefmark{3}}
\IEEEauthorblockA{\IEEEauthorrefmark{1}Independent Researcher, Abidjan, Côte d'Ivoire}
\IEEEauthorblockA{\IEEEauthorrefmark{2}Carnegie Mellon University Africa, Kigali, Rwanda}
\IEEEauthorblockA{\IEEEauthorrefmark{3}Carnegie Mellon University, Pittsburgh, PA, USA}
\IEEEauthorblockA{\IEEEauthorrefmark{4}University of Washington, Seattle, USA}
\thanks{Each author contributed ideas and/or writing to the paper. However, being an author does not imply agreement with every claim made in the paper, nor does it represent an endorsement from any author's respective organization.}
\thanks{This work was not supported by any organization. Correspondence to: Cédric Manouan (email:cmanouan@alumni.cmu.edu).}

\thanks{\noindent\textit{Preprint}. Copyright 2025 by the author(s).}
}

\input{map/World}

\let\oldtwocolumn\twocolumn
\renewcommand\twocolumn[1][]{%
    \oldtwocolumn[{#1}{
    \begin{center}
            \url{https://tinyurl.com/compute-tracker-for-africa}
            \includegraphics[width=0.58\textwidth]{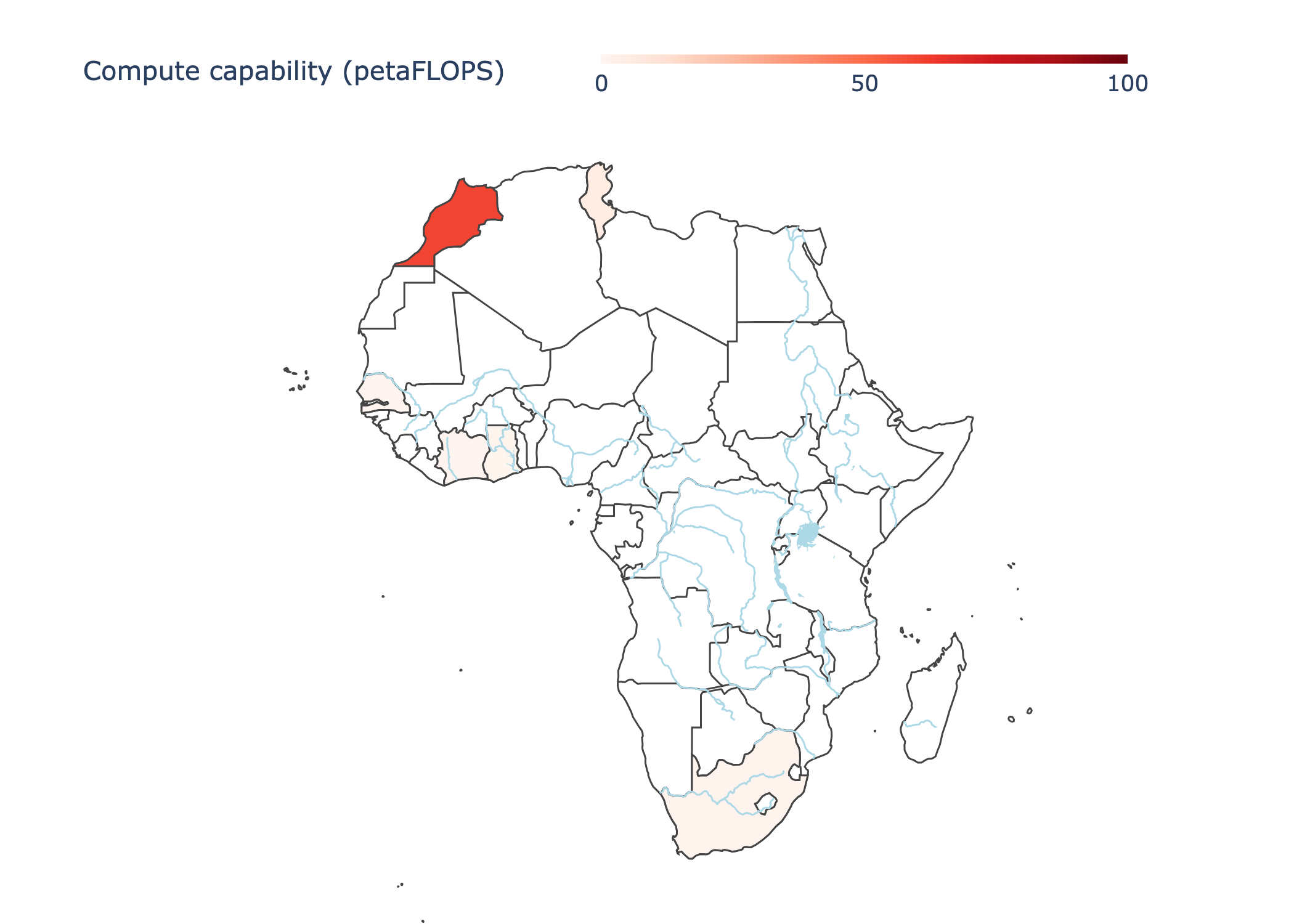}
            \captionof{figure}{Africa Compute Tracker (ACT). \textit{Status by September 2025}}
            \label{fig:compute_tracker}
        \end{center}
    }]
}

\DeclareRobustCommand{\IEEEauthorrefmark}[1]{\smash{\textsuperscript{\footnotesize #1}}}

\begin{document}

\maketitle
\begin{abstract}

Africa's participation in modern AI development is constrained by severe infrastructural and policy gaps. 
Important barriers include limited access to high-performance computing (HPC), restricted cloud access due to payment system mismatches, volatile exchange rates, and strict data sovereignty laws that fragment regional collaboration between African countries. 
Although initiatives such as Cassava AI's network of AI factories to be deployed across the continent signal the growing interest in adopting AI in Africa, these projects remain very targeted, while continental adoption still requires better coordination between African countries. 
Drawing on official African declarations on AI adoption across the continent, this paper offers both qualitative and quantitative evidence that sustainable AI adoption requires robust digital foundations through balanced access to compute, data, and the energy that makes it possible. 
We refer to these foundations as the ``right enablers'', considering them as crucial components for successful and sustainable AI adoption in Africa within the context of the global AI race.
We also introduce the \textit{Africa AI Compute Tracker (ACT)}, an interactive map to monitor the availability of AI-ready HPC systems throughout the continent. 
This tool represents the first open-source effort to consolidate data on Africa's evolving HPC landscape, and aims to encourage more transparency from local AI stakeholders while facilitating broader access for AI developers.
The work presented in this paper underscores the urgency of tangible actions aimed at closing the AI divide and allowing Africa to actively shape its AI future.
\end{abstract}



\section{Introduction}
Artificial intelligence (AI) is rapidly transforming industries, economies, and societies across the globe. 
From healthcare and agriculture to finance and education, AI systems offer unprecedented opportunities for automation, decision-making, and innovation \cite{RASHID2024100277}. 
However, access to these capabilities is unevenly distributed. 
Whereas developed regions benefit from abundant computing clusters and AI-optimized cloud platforms equipped with the latest accelerator hardware, Africa seems to have no other choice but to contend with resources that often fall short of the demands of modern AI workloads and are several generations behind state-of-the-art technologies \cite{tsado2024compute}. 
Although Africa boasts a youthful and burgeoning population, dynamic tech sectors, and budding artificial intelligence (AI) communities, it is notably underrepresented in the top tier of AI advancement. 
This is primarily due to the fact that the continent contributes less than 5\% to the global pool of research outputs \cite{Fonn2018Repositioning, thondhlana2021repositioning}. Despite these challenges, Africa's potential in the field of AI remains significant and ripe for exploration. \\
The continent's lateness to the \textit{global AI race} does not stem from a lack of talent or ambition, but from a structural digital divide rooted in limited \textit{compute} infrastructure, restricted access to global cloud platforms, and systemic barriers to technological scalability. 
The implications are far-reaching: without the means, i.e., the \textit{right enablers} to build and deploy AI systems, African countries risk deepening their dependency on foreign platforms, missing out on economic gains and limiting their ability to address uniquely regional problems with locally-informed solutions.

In this paper, we argue that addressing the \textit{AI divide} from a African perspective requires a concerted effort to expand access to foundational resources encompassing (but not limited to) broadband internet connectivity, AI-ready edge devices, and HPC systems. 
We also discuss the key barriers that limit AI adoption in Sub-Saharan Africa (SSA) and explore strategies to enable the adoption of inclusive AI in Africa under current conditions. 
Considering an African context more and more dominated by debates around digital sovereignty, we also argue that effective AI participation for African stakeholders hinges on coordinated interventions across critical dimensions including regulatory harmonization, integrated financial systems, data standards, and energy infrastructure among others. 
Using a policy synthesis approach, we draw on publicly available second-order data on infrastructure development, emerging continental strategies, and analogies from high-performing global models to identify high-leverage interventions. 
Furthermore, our analysis draws on insights from \textit{neural scaling laws} \cite{kaplan2020scaling, radford2022robustspeechrecognitionlargescale} supporting the idea that access to adequate computing capacity and data is central to effective AI development.

\textbf{\textit{Our contributions}}. Our main contributions are threefold: first, we introduce a taxonomy of barriers to AI adoption in sub-Saharan Africa, categorizing them into infrastructure, accessibility, governance, and human capital challenges. 
Second, we present the Africa Compute Tracker (ACT), a novel monitoring tool designed to map high-performance computing (HPC) systems and AI-ready infrastructure across the continent, enabling data-driven policy decisions and collaborative efforts to address resource disparities. 
Third, we propose an African AI infrastructure model structured across three stages (short-term, medium-term, and long-term) emphasizing a \textit{pragmatic hybrid approach} that aims to integrate on-premises and cloud-based solutions to balance scalability, cost-efficiency, and local sovereignty.

Starting with section \ref{taxonomy_of_barriers}, we contextualize our analysis by categorizing the different sets of barriers preventing SSA from fully leveraging AI. 
In Section \ref{hpc_and_trends}, we transition to an analysis of AI compute infrastructure encompassing (1) an assessment of HPC systems currently deployed in SSA and emerging initiatives to establish AI-ready facilities, (2) the Africa Compute Tracker (ACT) detailing its methodology, data sources, and collaborative framework to identify missing HPC installations, and (3) a comparison of domestic AI factory buildouts with hybrid models integrating rented and reserved cloud computing platforms.
Section \ref{africas_path_forward} describes our proposed approach to fostering inclusive AI development in SSA, covering capacity building initiatives, more long-term efforts to support local AI innovators, and a sustainable model for AI compute in SSA.
Limitations of this article and future work are discussed in section \ref{limitations}.

\section{Taxonomy of Barriers to AI Adoption in SSA}
\label{taxonomy_of_barriers}
Over the last decade, the global AI research and development landscape has been shaped by the widespread adoption of accelerator hardware, namely \textit{graphics processing units} (GPUs), which became essential with the rise of deep learning.\footnote{Deep learning is an approach to machine learning that uses (multi-layered) neural networks to automatically learn representations and patterns from large amounts of data.} 
Their parallel processing capabilities made them particularly suitable for training large AI models, driving breakthroughs in fields such as computer vision \cite{yolov12016redmon}, natural language processing (NLP) \cite{vaswani2023attentionneed}, automatic speech recognition \cite{radford2022robustspeechrecognitionlargescale}, or even computational biology \cite{Jumper2021}. 
Thus, access to high-performing GPUs has become a key determinant of who can participate in modern AI development \cite{sastry2024computingpowergovernanceartificial}.

Table \ref{tab:taxonomy_of_barriers} presents a comprehensive taxonomy of the key barriers to AI adoption in Sub-Saharan Africa, categorizing them into dimensions of infrastructure, accessibility, governance, and specialized human capital.

\begin{table*}[ht]
\centering
\caption{A taxonomy of key barriers to AI adoption in SSA}
\small
\begin{tabular}{@{}p{4cm} p{12cm}}
\toprule
\textbf{Barrier} & \textbf{Levels}\\
\midrule
Infrastructure & Global AI chip shipments; US-imposed restrictions; high energy requirements \\
\midrule
Accessibility & Limitations of payment methods; Currency and pricing issues; Challenges in latency\\
\midrule
Governance & Fragmented AI policy landscape; Lack of coordination on regional (data) frameworks\\
\midrule
Specialized Human Capital & Shortage in STEM-educated individuals\\
\bottomrule
\end{tabular}
\label{tab:taxonomy_of_barriers}
\end{table*}

\subsection{Infrastructure Barriers}
\label{infra_barriers}

\subsubsection{Geopolitics of Global AI Chip Shipments}
\label{geopolictics_of_chips}

Nvidia\footnote{Nvidia is a technology company headquartered in Santa Clara, California, that is the global leader in accelerated computing.} GPUs have become the de facto AI accelerator \cite{sastry2024computingpowergovernanceartificial, mishra2023artificial} due to their strong integration with the deep learning ecosystem from the outset \cite{alexnet2012, cireşan2012multicolumndeepneuralnetworks, rajat2009unsupervised}. 
These specialized chips come in two main categories: 1) data center and workstation GPUs, which are optimized for professional applications, and 2) consumer GPUs designed primarily for gaming. 
Recent architectures namely \textit{Fermi} (2010) \cite{Nvidiafermi2010},  \textit{Kepler} (2012) \cite{Nvidiakepler2012}, \textit{Maxwell} (2014) \cite{Nvidiamaxwell2014}, \textit{Pascal} (2016) \cite{Nvidiapascal2016}, \textit{Volta} (2017) \cite{Nvidiavolta2016}, \textit{Turing} (2018) \cite{Nvidiaturing2018}, \textit{Ampere} (2020) \cite{Nvidiaampere2020}, \textit{Ada Lovelace} and \textit{Hopper} (2022) \cite{Nvidiah1002023}, as well as \textit{Blackwell} (2024-2025) \cite{Nvidiablackwell2025} have played a pivotal role in driving significant AI breakthroughs in this decade of rapid technological growth. Hence, Nvidia stands out as the main player in the AI accelerator field, securing a market share of about 94\% in Q1 2025, while the remaining share belongs to Advanced Micro Devices (AMD)\footnote{AMD is a prominent American multinational semiconductor company which is a major competitor to Intel and Nvidia. It is specializing in designing high-performance computing and visualization products including CPUs, GPUs, field-programmable gate arrays (FPGAs), and AI accelerators for data centers, gaming, and personal computers.} \cite{jpr2025GPU}. 
 
The state of AI report (compute index) 2024 \cite{stateofaireport2025} shares details about the evolution of AI compute platforms at the global scale. 
The report highlights the concentration of advanced Nvidia GPUs in a handful of national and private infrastructure hubs, with very little to no notable representation from Africa. 
Organizations like \textit{Meta}, \textit{Tesla}, and \textit{DeepSeek} control tens of thousands (often hundreds of thousands) of these chips, while the same resources seem far beyond the reach of many African institutions.
Although notable initiatives are emerging on the continent, one typically has to look several generations back to see more of Africa's GPU assets. 
With clear disparities between sub-regions, Southern Africa is considered a part of the \textit{compute South} (countries with presence of AI-ready compute) while most of the continent is part of the \textit{compute desert}, i.e., countries where public cloud computing resources are scarcely available \cite{Lehdonvirta_Wú_Hawkins_2024}.

The rapid adoption of accelerator hardware worldwide highlights the urgency for more equitable access to scalable, efficient, and sustainable digital infrastructure, especially in the global south. 
The presence of large-scale cloud service providers, i.e. \textit{hyperscalers}, on the continent is predominantly concentrated in the southern region, with South Africa hosting data center facilities for companies such as Microsoft, Amazon, Google, Oracle, Huawei, and IBM. 
These data centers primarily serve domestic enterprises, offering services including AI and machine learning capabilities. 
Nevertheless, the availability of cutting-edge GPUs in these data centers remains limited, less documented, or arguably nonexistent \cite{Lehdonvirta_Wú_Hawkins_2024}. 
Consequently, this restricts the ability for local AI practitioners to build or deploy large-scale AI models unless they use overseas data centers, introducing security, latency, and accessibility challenges.

\subsubsection{The Power Dynamics}
\label{power_dynamics}

\begin{figure}[ht]
    \includegraphics[width=0.5\textwidth]{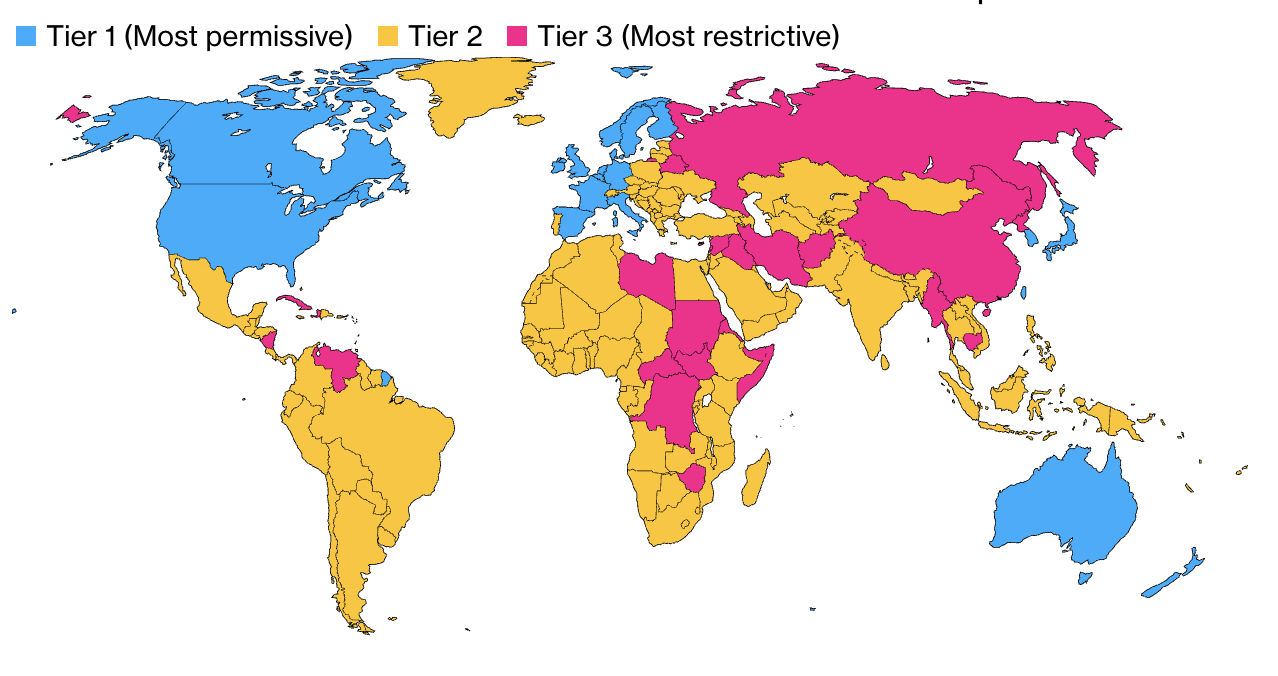}
    \caption{US-imposed curbs on global AI chip shipments. Source: Bloomberg Reporting \cite{us_imposed_curbs}}
    \label{fig:chip_restrictions}
\end{figure}

Starting in 2022, the United States (US) imposed restrictions on Nvidia AI chip exports, with most countries—especially in Africa, the Middle East, and parts of Asia—falling into Tier 2 or Tier 3, the latter being the most restrictive category (see Figure \ref{fig:chip_restrictions}). 
These new policies were designed to limit the global proliferation of advanced AI hardware by imposing tighter controls on US-designed GPUs. 
The decision was reportedly made to prevent China from accessing state-of-the-art AI compute that could enhance its internal capabilities, e.g., in the military domain. 
Nonetheless, one can argue that this step has inadvertently motivated the innovation behind \textit{DeepSeek}, one of China's frontier AI models.
Besides model architectures, \textit{DeepSeek} introduced several innovations in hardware-aware optimization to achieve reduced memory usage, processing efficiency, and therefore lower costs for training and serving large AI models.\\
For Africa, where access to advanced AI hardware is limited, such restrictions further exacerbate the persistent digital divide \cite{tbi2023digitaldivide}.
Without carve-outs or special agreements, many African nations are likely to find difficulties in acquiring the compute resources essential for modern AI research, development, and data center growth on their soil.

As African policymakers become increasingly aware of the importance of computing power across the continent, it should be noted that the idea of AI compute encompasses more than technical access. 
This concept is tightly connected to those of accessibility (section \ref{accessibility_barriers}), reliability and environmental sustainability (section \ref{energy_demand}), data security, and data sovereignty (section \ref{data_sovereignty}).\\
In many developing regions, the rather limited digital infrastructure restricts access to cloud and HPC resources. 
Notable efforts to accelerate the continent's digital transformation are underway, although core barriers still prevent Africa from fully harnessing its potential in the AI supply chains \cite{hopkins2025aisupplychainsemerging}.

\subsubsection{Energy Requirements}
\label{energy_demand}
The reliability and scale of electricity supply constitute a fundamental bottleneck in the deployment of HPC systems in SSA. 
AI-ready data centers and supercomputers (now mostly referred to as \textit{AI factories}) are not only capital intensive investments, but also extremely power-hungry. 
The yearly electrical consumption of a standard data center can be equivalent to that of anywhere between 25K and 100K households \cite{miyuru2016dataCenter, IEA2025}. 
And remarkably, some of those currently under construction on the continent are projected to use energy levels twenty times higher than this average consumption, requiring up to several megawatts \cite{IEA2025}.\\
Unfortunately, national grids in many SSA countries are currently unable to consistently provide the required level of electricity needed to power up the continent's AI ambitions. 
Although this problem is not uniquely African, SSA has particularly acute constraints. 
According to Afrobarometer,\footnote{Afrobarometer is a pan-African, research network with regional and national partners throughout Africa.} less than half of Africans use electricity supplied by a reliable grid connection \cite{lee2022reliableElectricity} and rural areas are especially disadvantaged. 
Almost a quarter of all African households use off-grid solutions, mainly solar panels and diesel generators \cite{msafiri2024energyGaps}.
Load-shedding, old transmission infrastructure, and a high cost of off-grid back-up diesel power make it unlikely that significant AI infrastructure will be sustainable.\\
Data centers globally accounted for roughly 1.5\% of electricity usage (415 terawatt-hours) in 2024, which is growing at more than four times the overall growth of electricity demand. 
In particular, AI data centers are becoming similar to industrial operations such as aluminum smelters, often concentrated in specific areas \cite{IEA2025}.
For example, \cite{IEA2025} also emphasized that nearly half of all data center capacity in the US is located in only five regional clusters which may facilitate the planning and delivery of energy. 
In comparison, there are no such highly concentrated regional clusters in SSA.
Instead, the region's digital landscape is dominated by a mix of private providers and state-owned/small-scale facilities like those presented in section \ref{hpc_in_africa}.

The International Energy Agency (IEA) is projecting that data centers will be responsible for roughly 10\% of the upcoming demand growth for electricity across the world by 2030, with regional variances. 
In developing economies like SSA, it is projected that up to 5\% of the new electricity demand will come from data centers, while the overall electricity demand for such economies continues to grow rapidly as they electrify \cite{IEA2025}. 
This situation adds another layer of pressure to energy systems that are already strained for reliability and affordability.

The AI-energy nexus in Africa, therefore, is nuanced.
On the one hand, building competitively impactful AI infrastructure requires continuous and substantial energy, which would further strain numerous (already fragile) grids and increase energy demand.
On the other hand, the push for AI offers an opportunity to rethink the power infrastructure altogether. 
Countries like Kenya, South Africa, Rwanda, and Senegal currently striving for nationwide electrification, are already leading renewable energy transitions \cite{ESMAP2025} and could be considered as pilot sites for integrated and clean energy-powered AI hubs or enclaves.\\
With specific SSA regions strategically developed, HPC infrastructure could serve as both a digital enabler as well an anchor load to entice investment into reliable and renewable energy systems. 
Meeting these broader needs requires collaborative development on infrastructure and policy, and coordinated alignment with national energy plans and digital transformation aspirations. 
Without doing so, the region risks missing the opportunity to connect to the computational backbone of the AI era, not due to lack of innovation, but because quite literally, the lights will not stay on.

\subsection{Accessibility Barriers}
\label{accessibility_barriers}

\subsubsection{Payment Method Limitations}
\label{payment_limitations}
Cloud service providers inherently require payment methods such as internationally accepted bank-backed cards. 
Though, according to the 2021 World Bank \textit{Global Findex Database} \cite{demirguc2022global}, credit card penetration in most African countries remained low, with sub-Saharan Africa averaging 4.2\%.  
In contrast, mobile money\footnote{Mobile money is a digital financial service that allows users to store, send, and receive money using a mobile phone without needing a traditional bank account.} is the dominant and widely preferred medium for personal and business transactions in many African countries due to its accessibility and deep integration into the daily life of millions of people \cite{demirguc2022global, gsma2025momo, klapper2025globalFindexDB}. 
But despite its widespread adoption across the continent, mobile money is barely supported by cloud platforms, except for some local ones like ST Digital.\footnote{ST Digital is a pan-African company offering digital transformation and cloud services, with presence in more than five countries across Africa.} 
This mismatch is a significant constraint, especially for local startups, researchers, and small businesses that lack access to traditional banking services or international credit lines. 
In fact, this situation is common in Africa given the prevalence of informal businesses with low digital use \cite{cruz2024africanBusinesses}.
As a result, a large segment of potential cloud users across the continent remains excluded from full participation in the digital economy, while already contributing to the overall economy with its approximately 85\% of informal workers \cite{ilo2023employment}.
This is arguably an untapped opportunity to improve digital services and promote economic growth on the African continent.

\subsubsection{Currency and Pricing Issues}
\label{currency_and_pricing_issues}

Globally accessible cloud services are predominantly priced in US dollars, particularly exposing users in low- and middle-income countries (LMICs) to volatile exchange rates and currency conversion fees. 
In countries with unstable currencies, this can make the effective cost of cloud access prohibitively high. For example, in Zimbabwe and Sudan, currency instability and inflation have severely eroded purchasing power.\footnote{The root causes of the currency instability differ for both countries. 
In Zimbabwe, these issues stem from a history of macroeconomic instability and frequent currency crises, while Sudan's economic collapse is largely due to the ongoing war, which has paralyzed production and increased reliance on mineral revenues.} 
In Nigeria and Ghana, frequent devaluations of the Naira and Cedi, respectively, have increased the local cost of dollar-denominated digital services. 
Naira devaluation has negatively affected small and medium-sized enterprises (SMEs) in the short term by increasing costs for imported inputs, making it harder to access foreign exchange and reducing consumer purchasing power. 
And even though local SMEs may become more competitive exporters due to the same situation, this devaluation can potentially have a negative impact on Nigeria's overall economic growth in the long run \cite{eromosele2025NGRcurrency,bakari2024NGRdevaluation}.
Thus, when converting the prices for cloud-based compute into African currencies, users from the region may face significantly increased costs.\\
These economic conditions contribute to the overall difficulty for African AI practitioners, a population mostly comprising students and young innovators (see section \ref{africa_ai_talents}), to consistently access relevant cloud platforms, thereby widening the digital divide.

\subsubsection{Challenges in Latency}
\label{latency_issues}
The limited presence of \textit{hyperscalers} with facilities located on the continent implies that data often has to travel long distances, leading to increased latency and adversely affecting the performance of latency-sensitive applications \cite{gsma2024compute}. 
In addition, reliable internet is crucial for efficient use of cloud services as it enables seamless data transfer, real-time processing, and smooth operation of applications hosted in the cloud. 
Hence a stable and fast internet connection is required for minimal latency, reducing the potential for delays in service delivery.
On the other hand, excess dependence on data centers located outside the continent increases Africa's exposure to latency issues and increases its vulnerability to service interruptions and external network failures.\\
Most of the time, African users are routed to cloud endpoints located in North America or Europe, bypassing the continent entirely and increasing network inefficiencies. 
Study by Babasanmi and Chavula \cite{babasanmi2022measuring} found that accessing servers from Amazon Web Services (AWS) or Azure cloud regions from Africa incurs median round-trip times (RTTs) of 74 to 84 milliseconds, compared to 12 to 15 ms typically experienced in Europe and North America.\\ 
Moreover, intra-African routing remains inefficient as more than 38\% of first-hop internet connections from African probes exit the continent, and nearly 50\% of inter-African traffic is routed through non-African exchange points.
This creates arguably avoidable bottlenecks and exposes the region to greater risks from global network outages \cite{chavula2017latency}. 
Furthermore in the context of speech/audio-based use cases, which are gaining traction in African AI, long distances between users and servers can lead to slower response times or delayed content loading \cite{Lehdonvirta_Wú_Hawkins_2024}.\\
In contrast, using content delivery networks (CDNs) with local points of presence (PoPs) in Africa reduces latencies to 29-65 ms, indicating an overall improvement of 25-87\% \cite{babasanmi2022measuring}.\\
The measurable gaps in service latency underscore the urgency of bringing cloud infrastructure closer to African users to ensure equitable access to HPC resources, and scalability of AI-driven applications.

\subsection{Governance Barriers}
\label{governance_barriers}

\subsubsection{The African AI Governance and Policy Landscape}
\label{ai_governance}

Leaders around the world have recognized the need for global AI governance, as well as an improvement in global cooperation \cite{UN2024GoverningAI}. 
Along the same lines, the AI governance landscape in Africa is evolving to address unique regional challenges while leveraging local opportunities. 
As pointed out by \cite{adams2024WhatisAIgov}, it involves establishing comprehensive frameworks that encompass laws, strategies, and policies tailored to African needs.
Key priorities revolve around 1) ensuring equitable benefits from AI advancements, 2) protecting human rights, and 3) safeguarding the environment amid rapid technological growth. 
Governance efforts also focus on creating robust ecosystems for AI development, preventing exploitation of resources, and fostering accountability in AI deployment.

Considering that approximately 30\% of AU member states have published validated or draft AI strategies by September 2025 \cite{africaDPA2025} (with over 20 more actively developing theirs), the growing recognition of AI's strategic importance to African leaders cannot be neglected. 
This proactive approach demonstrates the commitment of numerous governments to harness the potential of this technology for economic growth, innovation, and social advancement. 
Yet, the existence of these strategies alone is not a guarantee of success. 
While the prioritization of stand-alone strategies is promising, effective implementation definitely hinges on the existence of a comprehensive ecosystem. 
Such an ecosystem should be grounded in essential foundational activities like capacity building, fostering collaboration among all stakeholders, providing access to cutting-edge technologies, ensuring sustainable funding models, and establishing robust digital infrastructure \cite{WorldBank2025_AIHandbook}. 
These elements are crucial for fully realizing the potential of these strategies across various domains.
Unless substantial commitments in these areas are made, African AI strategies will remain aspirational documents rather than catalysts for actual progress.

African nations face considerable obstacles as they establish AI policy frameworks and strategies on top of unfulfilled digitalization ambitions. 
A critical concern is the excessive focus on integrating AI across various economic sectors without taking into account how other digital technologies could strengthen the intended socioeconomic objectives \cite{abdella2025racedStrategies}.
Moreover, while some countries have adopted multi-stakeholder engagement, the limited representation of diverse groups in the formation of their AI policy frameworks may exacerbate digital disparities within the continent.\\
Lastly, attempting to build AI foundations by simultaneously establishing digital infrastructure and demonstrating immediate impact often leads to prioritization of short-term (but more visible) outcomes at the expense of long-term (but more impactful) actions. 
Going down that route exposes to risks of creating a cycle of fragmented efforts where initial AI projects might produce immediate benefits, but ultimately fail to properly scale due to an underdeveloped basis. 
Unfortunately, the pressure to showcase rapid progress can inadvertently undermine the very foundations needed for lasting success.

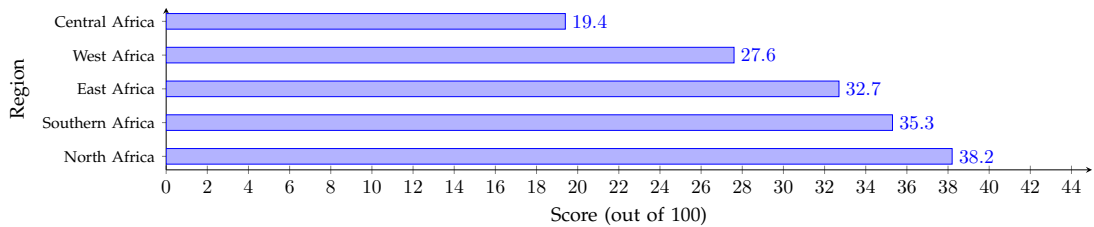
\begin{figure*}[!ht]
\centering

\resizebox{.8\textwidth}{!}{
\begin{tikzpicture}
\begin{axis}[
    xbar,
    width=1.0\textwidth,
    height=4.5cm,
    xlabel={Score (out of 100)},
    ylabel={Region},
    ytick=data,
    axis lines=left,
    yticklabel style={font=\footnotesize, align=center, anchor=east},
    symbolic y coords={
        North Africa, 
        Southern Africa, 
        East Africa, 
        West Africa, 
        Central Africa,
    },
    yticklabel style={
          /pgf/number format/.cd,
         use comma=false, 
        1000 sep ={}
    },
    nodes near coords,
    xmin=0, 
    xmax=45,
    bar width=8pt,
    enlarge y limits=0.1,
    title={}
]
\addplot coordinates {
    (38.2,North Africa)
    (35.3,Southern Africa)
    (32.7,East Africa)
    (27.6,West Africa)
    (19.4,Central Africa)
};

\end{axis}
\end{tikzpicture}
}

 \caption{African AI Talent index (regional averages). Redrawn from \cite{qhalaAIreadiness2025}.}
\label{fig:ai_talent_index}
\end{figure*}

\subsubsection{Data Governance Frameworks}
\label{data_sovereignty}
Policies on data sovereignty, although essential for protecting national interests and citizen privacy, can impose significant constraints on local AI development, particularly in emerging markets like those in SSA. 
These policies often require that data generated within a country be stored, processed, and sometimes even analyzed within national borders. 
In essence, this protects against foreign exploitation and ensures regulatory control, but it can inadvertently result in limited access to global-scale digital platforms that offer the compute and tooling necessary for AI research and deployment worldwide.\\
On the one hand, the African Union (AU) is promoting cross-border data flows (CBDFs) with its data policy framework \cite{AfricanUnion2022DataPolicyFramework}. 
This framework emphasizes a comprehensive approach to data governance that extends beyond just personal data protection, which has been a primary focus worldwide and on the continent. 
More specifically, it is deeply connected with the development of regional and continental data infrastructure to support digital integration, research, development, and innovation in areas such as AI and big data analytics. \\
On the other hand, there seems to be a disconnect between AU's continental initiatives and priorities within member states \cite{fola2024AIpoliciesinAfrica} as exemplified by the low rate of ratification of the \textit{Malabo Convention},\footnote{Article 36 of the AU Convention on Cyber Security and Personal Data Protection (commonly referred to as the Malabo Convention) states that the treaty would come into effect once 15 ratifications were achieved.} which only entered into force in 2023 after being ratified by Mauritania almost ten years after its adoption. 
This relative reluctance can be attributed to several factors.\\
Primarily, data policy is often seen as a matter deeply tied to national sovereignty, with countries prioritizing their own regulatory frameworks over supranational agreements \cite{musoni2024cross}. 
This perspective is evident in the varied approaches to data protection and localization across African countries like Senegal, Nigeria, Zambia, Côte d'Ivoire, Benin, and Morocco, where some enforce strict mandates that require substantial policy adjustments and investment to align with AU's goal of harmonized cross-border data flows. 
This situation is also the consequence of differences in levels of digital readiness, which influences countries' willingness or ability to commit to such conventions.\\
Secondly, political considerations and the complexities involved in coordinating multiple stakeholders across different jurisdictions contribute to the slow ratification process \cite{musoni2024cross, Ndemo2023}. 
In fact, intergovernmental collaboration involves the navigation of bureaucratic processes and institutional capacities that may differ significantly across the continent \cite{Yusuf2025}. 

Collectively, these factors highlight an obstacle that the AU and continental entities like \textit{The Smart Africa Alliance} are trying to overcome: reconciling diverse local priorities with overarching regional goals for data (and more generally digital) governance across the continent \cite{musoni2024cross}.
Considering that many African countries are still building their foundational digital infrastructure, a balanced approach to regulation, especially in the data domain, would be more appropriate. 
Too many restrictive policies and compliance burdens can deter \textit{hyperscalers} from offering services in countries with strict or ambiguous legal frameworks, thereby exacerbating the digital divide and isolating African regions due to substantial additional costs and operational complexities.

\subsection{Shortage in Specialized Human Capital}
\label{shortage_of_human_resources}

The AI talent readiness index\footnote{The index was developed by \textit{Qhala}, an African digital innovation and transformation company based in Nairobi, Kenya.} for Africa \cite{qhalaAIreadiness2025} highlights distinct regional patterns in the continent's capacity to effectively participate in AI supply chains as an actual contributor. 
The first pillar of this index assesses nations' digital skills capabilities. 
It rates the availability and quality of AI-related education, the volume of science, technology, engineering, and mathematics (STEM) graduates, and the overall AI literacy within the workforce. 
Key indicators include secondary school completion rates, the proportion of the labor force holding higher education degrees, the representation of women in STEM fields, the number of institutions offering AI/Machine Learning education, and the density of developers per million people.

Regional average scores are presented in Figure \ref{fig:ai_talent_index}.\\
North Africa leads AI adoption with Tunisia, Egypt, Algeria, and Morocco driving progress through advanced education, high developer density, and digital initiatives, while East Africa shows promise despite disparities. 
West Africa holds a relatively middle position, with Ghana and Nigeria showing potential, whereas Central Africa faces significant challenges requiring urgent infrastructure and governance improvements. 
South Africa leads the Southern region with strong digital infrastructure but struggles with regional disparities.

The shortage of tech-ready domestic professionals (science teachers and instructors, data center engineers, network architects, cloud operations specialists, etc.) is a critical constraint on the effective utilization of the region's existing digital infrastructure \cite{agcc2025compute}.
As an example, the lack of qualified science educators combined with less than 25\% of African higher education students pursuing STEM qualifications \cite{worldbank2023empowering,Mutsvangwa2021STEM} hinder the development of a pipeline of future tech professionals, perpetuating a cycle of dependency on foreign expertise to build local solutions. 
This skills gap is further worsened by brain drain, where skilled individuals leave the continent for opportunities in more developed economies like the US, the UK, Canada, and Australia among others \cite{nepad2021,ddp2025, adesote2018brain}.

\section{Compute: Current State, Emerging Initiatives, and Insights from Other Regions}
\label{hpc_and_trends}

\subsection{AI infrastructure in Africa}
\label{ai_infra_in_africa}

\subsubsection{The Landscape of High-performance Computing}
\label{hpc_in_africa}
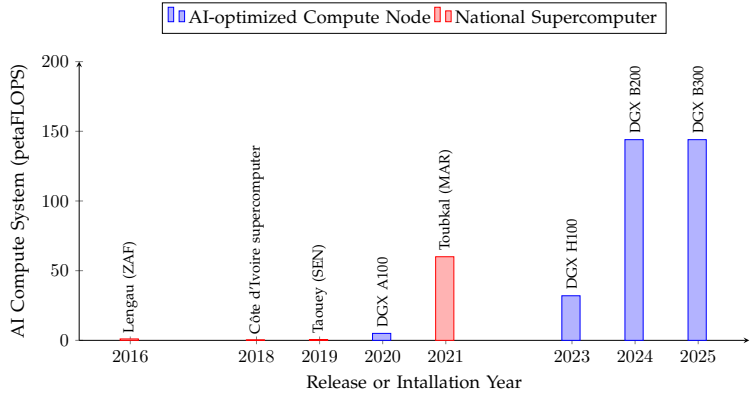
\begin{figure*}[!ht]
\centering

\resizebox{0.55\textwidth}{!}{
\begin{tikzpicture} 
\begin{axis} [
    ybar,
    width=0.8\textwidth,
    height=7cm,
    axis lines=left,
    xlabel={Release or Intallation Year},
    ylabel={AI Compute System (petaFLOPS)},
    xmin=2016, 
    xmax=2025,
    ymin=0, 
    ymax=200,
    xtick={
        2016,
        2018,
        2019,
        2020,
        2021,
        2023,
        2024,
        2025
    },
     xticklabel style={
          /pgf/number format/.cd,
         use comma=false, 
        1000 sep ={}
     },
    enlarge x limits=0.09,
    yticklabel style={/pgflots/number format/.cd},
    legend style={at={(0.5,1.1)}, anchor=south, legend columns=-1},
]
\addplot+[ybar, bar shift=-0.02cm] coordinates {
    (2020,5.0)
    (2023,32.0)
    (2024,144.0)
    (2025,144.0)

};
\addplot+[ybar, bar shift=-0.02cm] coordinates {
    (2016,1.029)
    (2018,0.3226)
    (2019,0.5376)
    (2021,60)
};

\node[anchor=west, rotate=90, align=center, font=\footnotesize] at (axis cs:2024,144.0) {DGX B200}; 
\node[anchor=west, rotate=90, align=center, font=\footnotesize] at (axis cs:2023,32.0) {DGX H100}; 
\node[anchor=west, rotate=90, align=center, font=\footnotesize] at (axis cs:2020,5.0) {DGX A100}; 
\node[anchor=west, rotate=90, align=center, font=\footnotesize] at (axis cs:2021,60) {Toubkal (MAR)}; 
\node[anchor=west, rotate=90, align=center, font=\footnotesize] at (axis cs:2019,0.5376) {Taouey (SEN)}; 
\node[anchor=west, rotate=90, align=center, font=\footnotesize] at (axis cs:2018,0.3226) {Côte d'Ivoire supercomputer}; 
\node[anchor=west, rotate=90, align=center, font=\footnotesize] at (axis cs:2016,1.029) {Lengau (ZAF)}; 
\node[anchor=west, rotate=90, align=center, font=\footnotesize] at (axis cs:2025,144.0) {DGX B300}; 

\legend{AI-optimized Compute Node, National Supercomputer}
\end{axis}
\end{tikzpicture}
}

 \caption{African supercomputers versus state-of-the-art DGX systems.}
\label{fig:african_ai_hpc_vs_sota}
\end{figure*}

Africa hosts up to 35 supercomputing systems across 11 countries \cite{hpcsystems2024} aimed at boosting research and development (see table \ref{tab:hpc_africa} for a summary of recent installations). 
Meanwhile, when comparing Africa's HPC landscape with Nvidia's state-of-the-art GPU offerings, the gap in compute capability becomes both clear and concerning.\\
Current African HPC systems, though limited in scale, represent important progress in establishing accelerated and distributed computing platforms in the region. 
Their presence signals an interest in accelerated computing, even as most of the continent's compute capacity remains CPU-centric and largely focused on traditional scientific workloads, e.g. weather forecasting.
In the following lines, we explore these installations in a broader context.

Morocco's \textit{Toubkal}\footnote{Toubkal is the only African supercomputer listed in the TOP500 dataset of November 2024; see \url{top500.org/lists/top500/list/2024/11/}} supercomputer reportedly delivers 3.15 petaFLOPS.\footnote{This is the performance reported by the host institution before the subsequent upgrades were made. It is likely that the overall capacity might have improved to about 60 petaFLOPS as of March 2025.} 
This performance, in the context of modern AI, is arguably equivalent to deploying one Nvidia Deep GPU Xceleration (DGX) A100 node, capable of delivering about 5 petaFLOPS of performance. 
\textit{Toubkal} ranked 316th globally, before the system was upgraded with about 20 A100 and 12 H100 GPUs, respectively in March 2022 and March 2025 \cite{kissami2025toubkal, kissami2025toubkalv2}.

South Africa's \textit{Lengau} supercomputer is powered by 36 Tesla V100s and provides over 1 petaFLOP. 
It was ranked 121st on the world's TOP500 list of supercomputers at its launch by the center for high performance computing (CHPC). 
Though, in the current HPC landscape, this is below the performance of a single DGX-2 V100\footnote{The V100 was rapidly adopted because it introduced Nvidia tensor cores, units specifically designed to accelerate matrix operations particularly for deep learning and HPC.} node (2 petaFLOPS) released in early 2018.

Other notable GPU-accelerated supercomputers in SSA include Senegal's \textit{Taouey} and that of Côte d'Ivoire. 
\textit{Taouey}, powered by 48 V100s with an overall compute capability of 537.6 teraFLOPS, has reportedly supported applications in climate modeling, agriculture research, and AI through NLP for local languages.\\
Côte d'Ivoire's supercomputer is powered by 24 Kepler K80 GPUs, each offering up to 5.6 teraFLOPS of single-precision performance for a total compute capability of 322.56 teraFLOPS.\\
In contrast, Nvidia's DGX Blackwell B200 launched in 2024 delivers between 72 and 144 petaFLOPS per node, figures that eclipse the aforementioned HPC installations by multiple orders of magnitude (see Figure \ref{fig:african_ai_hpc_vs_sota}). 
This disparity reveals a missed opportunity for Africa, resulting in domestic innovators being compelled to rely on smaller, and sometimes less competitive AI models due to a lack of in-house computing power and insufficient funding to support growth ambitions. 
In fact, it is common for African AI practitioners to access AI compute resources through gifted cloud credits, as exemplified by the recent development of an open-source text-to-speech (TTS) model for Senegal's Wolof \cite{xTTS-v2-wolof} by \textit{GalsenAI}\footnote{\url{https://galsen.ai}} with the support of the startup \textit{Caytu}\footnote{\url{https://caytu.ai/}} which provided the required compute.
Such a situation can further impair SSA's involvement in the AI supply chains and places the region's AI workforce at a disadvantage by necessitating reliance on AI technologies developed and accessible mainly in \textit{GPU rich} regions \cite{tsado2024compute}.

\begin{table}[!ht]
\centering
\caption{GPU-powered HPC installations in Africa vs notable state-of-the-art AI-optimized compute systems(based on published data)}
\small
\begin{tabular}{@{}p{1.8cm} p{2.48cm} p{1.6cm} p{1.5cm}}
\toprule
\textbf{Country} & \textbf{System} & \textbf{Performance (petaFLOPS)} & \textbf{GPU Acc.}\\
\midrule
Tunisia & DGX A100 (2020) & 5.0 & A100 (x8) \\
Morocco & Toubkal (2020)\dag & 3.15 & None \\
South Africa  & Lengau (2016)\ddag & 1.029 & V100 (x36)\\
Senegal & Taouey (2019)\P & 0.5376 & V100 (x48)\\
Côte d'Ivoire  & Unnamed (2018)\S & 0.3226 & K80  (x24)\\
\midrule
N/A  & DGX H100 (2023) & \textbf{32.0} & H100 (\textbf{x8})\\
N/A  & DGX B200 (2024) & \textbf{72.0-144.0} & B200 (\textbf{x8})\\
N/A  & DGX B300 (2025) & \textbf{72.0-144.0} & B300 (\textbf{x8})\\
\bottomrule
\end{tabular}
\vspace{0.5em}

\raggedright
\scriptsize
\textsuperscript{\dag}\url{https://cc.um6p.ma/toubkal-super-computer} \\
\textsuperscript{\ddag}\url{https://wiki.chpc.ac.za/chpc:lengau} \\
\textsuperscript{\P}\url{https://cineri.sn/carasteriqtiques/}\\
\textsuperscript{\S}\url{https://cncci.edu.ci/cncci/}

\label{tab:hpc_africa}
\end{table}
In our analysis, we specifically consider the age and type of the accelerator hardware that powers HPC installations because for Africa to contribute to AI development, it is crucial to prioritize relevant hardware with cutting-edge features. 
The underlying motivations are simple: 
\begin{itemize}
    \item \textbf{Compatibility issues}: advanced techniques like \textit{flash attention} and its variants \cite{dao2022flashattention, dao2023flashattention2,shah2024flashattention3} rely on specialized hardware such as Ampere- or Hopper-based GPUs, which boast advanced processing units for \textit{mixed-precision}\footnote{Mixed precision is a technique that combines different numerical precision levels, most commonly single-precision (\(32\)-bit or FP32) and half-precision (\(16\)-bit or FP16), to speed up computations and reduce memory usage in deep learning.} computations, warp-level parallelism primitives, and featuring asynchronous \textit{tensor cores} or \textit{tensor memory accelerator} (TMA).
Many of these features available on recent GPUs are essential for performance optimizations. 
Older accelerators lack these capabilities, restricting their ability to fully exploit the algorithms' potential or even run them at all.

\item \textbf{Efficiency issues}: unlike \textit{flash attention}, \textit{DeepSeek}'s \textit{multi-head latent attention} (MHLA) technique \cite{deepseekai2024deepseekv2} can run on older GPUs like the Tesla V100, but it is significantly slower due to the GPU's limitations in VRAM\footnote{Video Random Access Memory (VRAM) is a dedicated type of memory on a GPU, used to store and quickly access graphics-related data.}, memory bandwidth, and tensor cores. This is a major concern, given that large language model (LLM)-based applications are gaining traction with use cases in key sectors like agriculture or healthcare.
And efficient deployment of LLMs is essentially made possible by leveraging the aforementioned techniques and many others that seem constrained by the available hardware.
\end{itemize}

\begin{figure*}[!ht]
\centering
\resizebox{1.3\columnwidth}{!}{
\begin{tikzpicture} 
\begin{axis}[
    width=\textwidth,
    height=7.5cm,
    axis lines=left,
    xlabel={Year},
    ylabel={Number of papers},
    xmin=2018, xmax=2024,
    ymin=0, ymax=18000,
    xtick={2018,2019,2020,2021,2022,2023,2024},
     xticklabel style={
          /pgf/number format/.cd,
         use comma=false, 
        1000 sep ={}
     },
    legend style={at={(0.5,1.1)}, anchor=south, legend columns=-1},
    smooth
]

\addplot+[mark=none, thick, orange,
    nodes near coords,
    every node near coord/.append style={
        font=\footnotesize,
        rotate=0,
    }
] coordinates {(2018,0)(2019,0)(2020,111)(2021,746)(2022,3080)(2023,8390)(2024,17400)};
\addlegendentry{A100}

\addplot+[mark=none, thick, pink] coordinates {(2018,0)(2019,0)(2020,199)(2021,1190)(2022,3440)(2023,6460)(2024,8160)};
\addlegendentry{RTX 3090}

\addplot+[mark=none, thick, cyan,
    nodes near coords,
    every node near coord/.append style={
        font=\footnotesize,
        rotate=0,
    }
] coordinates {(2018,353)(2019,1630)(2020,4060)(2021,6340)(2022,7390)(2023,7550)(2024,6770)};
\addlegendentry{V100}

\addplot+[mark=none, thick, yellow!70!black] coordinates {(2018,0)(2019,0)(2020,0)(2021,0)(2022,248)(2023,829)(2024,4400)};
\addlegendentry{4090}

\addplot+[mark=none, thick, purple,
] coordinates {(2018,388)(2019,910)(2020,2310)(2021,3050)(2022,3030)(2023,2970)(2024,3260)};
\addlegendentry{2080}

\addplot+[mark=none, thick, red!70!black] coordinates {(2018,0)(2019,0)(2020,0)(2021,0)(2022,31)(2023,160)(2024,3910)};
\addlegendentry{H100}

\addplot+[mark=none, thick, teal] coordinates {(2018,0)(2019,161)(2020,871)(2021,1110)(2022,1230)(2023,1300)(2024,982)};
\addlegendentry{Titan}

\addplot+[mark=none, thick, lime!60!black] coordinates {(2018,177)(2019,303)(2020,445)(2021,589)(2022,559)(2023,519)(2024,763)};
\addlegendentry{Jetson}

\addplot+[mark=none, thick, blue] coordinates {(2018,477)(2019,842)(2020,1160)(2021,1100)(2022,1000)(2023,753)(2024,629)};
\addlegendentry{P100}

\addplot+[mark=none, thick, black] coordinates {(2018,464)(2019,542)(2020,669)(2021,535)(2022,343)(2023,203)(2024,137)};
\addlegendentry{K80}

\end{axis}
\end{tikzpicture}
}
 \caption{Cited Nvidia GPU usage in open-source AI papers from 2018-2024 (estimates from the state of AI report/compute index and Zeta Alpha).}
\label{fig:Nvidia_citations}
\end{figure*}
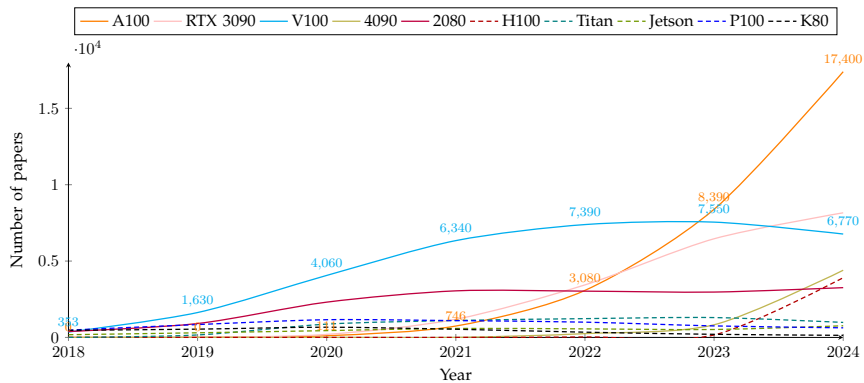

Figure~\ref{fig:Nvidia_citations} reinforces the centrality of specific AI hardware by tracking the increase in citations of specific Nvidia chips in AI-related publications. 
We see that starting from its release in 2020, the A100 has gained more popularity to rapidly establish itself as the most frequently cited accelerator by 2024. 
Following a similar trend, the Hopper H100 is quickly rising in prominence. 
It is also worth noting the decrease in citations for the Tesla V100 as the AI community gradually shifts toward the latest chips.
Meanwhile, the growing presence of the RTX 3090 and 4090\footnote{Nvidia RTX 3090s and 4090s (24GB VRAM each) are originally gaming GPUs that meet the requirements for relatively low-scale AI R\&D.} suggests that many researchers, particularly in constrained environments, are relying on consumer-grade accelerators for their tasks. 
This is mainly the case in academic and small business/startup settings where budget limitations necessitate the use of lower-grade hardware for R\&D activities.

\textbf{\textit{Implications for Local AI Talents}}. In an article co-authored with \textit{Zindi} and \textit{Alliance4AI}, the United Nations Development Programme (UNDP) argues that only 5\% of Africa's AI talent has access to the computational resources required to support their AI-related projects \cite{tsado2024compute}. 
This conclusion was drawn from an analysis of compute usage by up to 11K data scientists from the \textit{Zindi} community, which represents both Africa's largest network of AI practitioners and the continent's reference platform for AI-related competitions.
Furthermore, the remaining 95\% of AI builders on the platform rely on accelerators provided by Google's hosted cloud service \textit{Colab},\footnote{\url{https://colab.research.google.com/signup}} offering default access to GPUs like the Nvidia Kepler K80, Turing T4, or Pascal P100, free of charge for a couple of hours weekly or A100s on paid subscriptions. 
Consequently, the limited access experienced by African innovators implies that they may have to wait much longer to iterate on experiments, compared to a few minutes for their peers in developed countries \cite{tsado2024compute}, who have access to much faster AI compute.

\subsubsection{Emerging African Compute Initiatives}
\label{emerging_initiatives}

\textit{Cassava Technologies} (in collaboration with Nvidia) is building a network of AI factories across Africa. 
This project is estimated at about \$720 million and plans to deploy up to 12K GPUs\footnote{Details about the GPUs powering these AI factories have not been disclosed, but it is expected to be a mix of high-end data center units.} across multiple countries, including South Africa, Egypt, Nigeria, Kenya, and Morocco \cite{Nvidia2024cassava}.\\
A few months following \textit{Cassava AI}'s announcement, \textit{Synectics Technologies} and \textit{Schneider Electric} have also unveiled a partnership to establish Uganda's first AI factory within the 600MW \textit{Karuma} hydropower station.\footnote{\url{www.uegcl.com/power-plants/karuma-hydropower-project}} 
Referred to as the \textit{Aeonian Project}, this 100MW hybrid Tier-4 Plus (4+) off-grid facility\footnote{Typically, the installation will be a hybrid hyperscale data \& high-performance computing centre (DHPC).} is backed by global partners including Nvidia, the GIZ, HAUS, the EU Development Fund, and MDCS.AI among others. 
The project comprises multiple stages, with its initial phase—which encompasses a 15MW AI module and a 10MW sovereign supercomputer (USIO)—planned for launch in the second half of 2026 with full capacity expected by H2 2027.
USIO is expected to incorporate Nvidia's Blackwell GB300 GPU, along with a complete Nvidia enterprise AI ecosystem, designed to support diverse application areas like healthcare and life sciences among others.\footnote{\url{https://renewableenergynews.co.ke/synectics-technologies-and-schneider-electric-announce-africas-first-sovereign-supercomputer-and-ai-factory-ecosystem-the-aeonian-project/}} 
Notably, the \textit{Aeonian Project} is planned to operate solely on renewable energy, making it one of the world's greenest AI facilities.\\
Another key initiative led by Microsoft in partnership with the UAE-based AI firm G42 is the announcement of a \$1 billion investment to build a data center in Kenya, aimed at expanding Microsoft's Azure services to East Africa \cite{microsoft2024g42}.

The Africa Green Compute Coalition (AGCC), a collaborative initiative led by the UNDP, has been created to coordinate sustainable AI computing across Africa. 
As part of the AI Hub for Sustainable Development,\footnote{The AI hub for sustainable development is an initiative co-led by the UNDP and the Italian G7 presidency, see \url{www.aihubfordevelopment.org}.} AGCC advocates for a robust, resilient, and sustainable computing ecosystem by establishing a pan-African governance and financing framework. 
It seeks to unlock the potential of both cloud and on-premises GPU-as-a-service options. 
The interim findings of the AGCC report \cite{agcc2025compute} present an immediate investment opportunity worth \$150 million, addressing a documented demand for \textbf{7 million GPU hours} solely for AI model training.

These initiatives mark a significant step towards embedding Africa in global AI supply chains by addressing a key prerequisite: democratized access to AI compute. 
The location of HPC resources within African borders can stimulate regional innovation ecosystems, support the development of domestic AI startups, and foster public-private partnerships (PPPs) to deploy AI applications in priority sectors.\\
It would be fair to acknowledge that African countries are not equally prepared for AI adoption, with varying levels of prerequisites before full implementation can occur. 
However, by observing how AI compute projects are being rolled out, it seems there may be competing efforts emerging across different parts of the continent, similar to data-related initiatives (see section \ref{data_sovereignty}). 
Nonetheless, we also emphasize that regional coordination and national sovereignty in AI are not mutually exclusive but rather complementary processes.

\subsubsection{Monitoring Africa's AI Computing Power}
\label{act}

\begin{figure}[ht]
    \centering
    \includegraphics[width=0.33\textwidth]{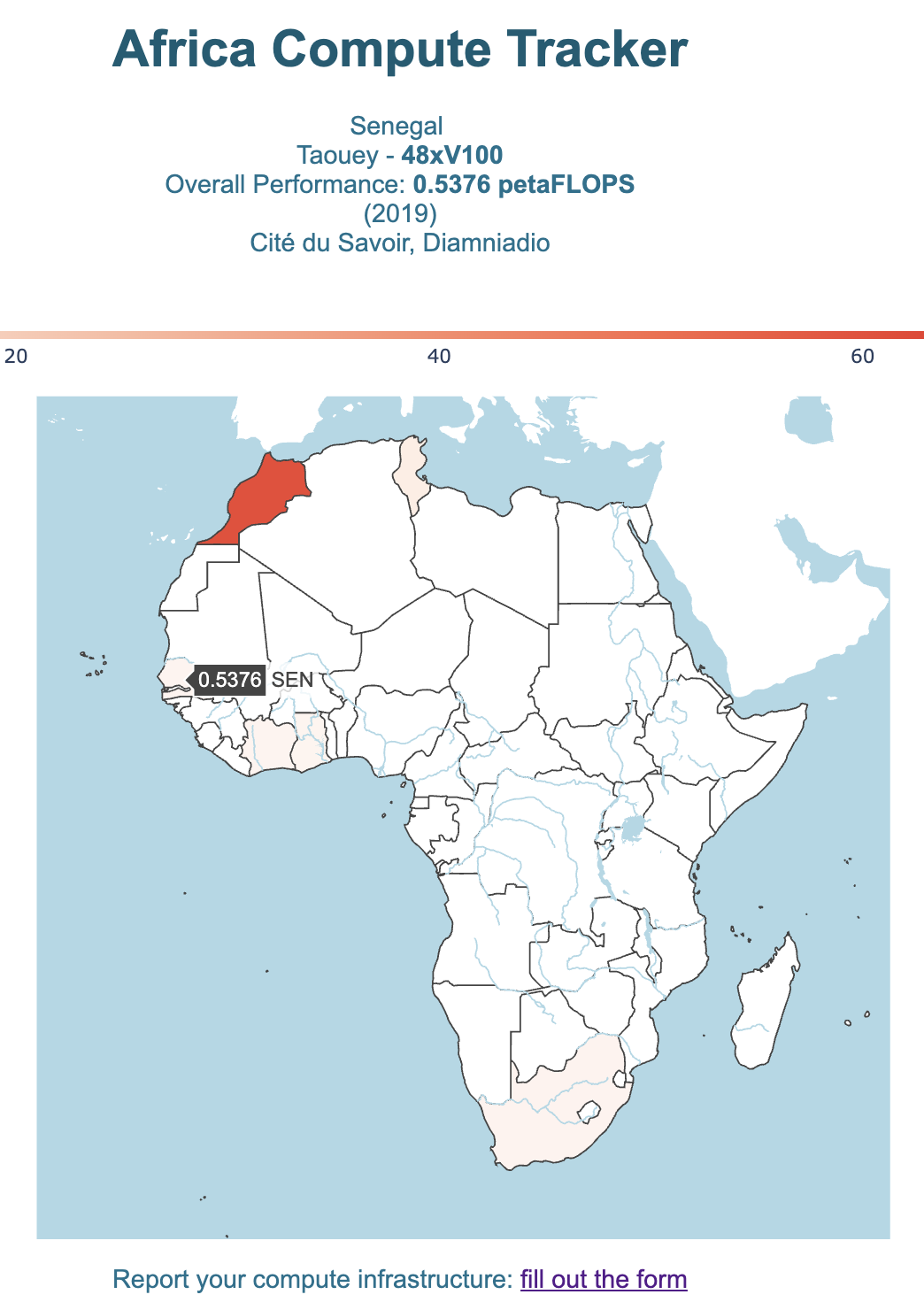}
    \caption{A Snapshot of the Africa Compute Tracker. Accessible from: \url{https://tinyurl.com/compute-tracker-for-africa}.}
    \label{fig:act}
\end{figure}

In this section, we present the Africa AI Compute Tracker (ACT), an interactive tool designed to provide transparent, data-driven insights into the distribution and capacity of live HPC and AI-ready infrastructure across Africa (see Figure \ref{fig:act}). 
Its primary purpose is to serve as a resource for targeted actions, e.g. investments, in compute resources, energy systems, and domestic cloud platforms, enabling policymakers, researchers, and stakeholders to address regional disparities in AI infrastructure.
This tool was built by aggregating publicly available data from diverse sources including manufacturer reports, academic publications, government records, and industry partnerships, to create a comprehensive map of HPC installations.\\
Modern AI supercomputers typically operate at the petaFLOP ($10^{15}$) and exaFLOP ($10^{18}$) scales. 
With a focus on systems with capabilities exceeding 0.2 petaFLOP, ACT ensures its data remains relevant in the broader context of AI infrastructure while including much smaller systems currently installed across Africa.
This orientation allows ACT to highlight critical gaps in AI-ready infrastructure, such as regions with limited access to HPC systems or those with installations that fall behind global trends in modern AI supercomputers \cite{pilz2025trendsaisupercomputers}.\\
More specifically, the map shows the highest GPU type available as well as the highest AI compute capacity at the country level measured in petaFLOP. \\
Recognizing the evolving nature of the global AI compute landscape, ACT includes a public form for submitting details about missing HPC systems, enabling continuous updates and community-driven data collection. 
This participatory approach hopes to guarantee that ACT remains current and fosters a sense of ownership among local stakeholders. 
We hope that ACT helps identify regions where short-term investments in AI compute facilities are most urgent, while contributing to tracking progress toward long-term goals like domestic cloud scalability and hybrid (on-premise+cloud) architectures.

\subsection{Scaling AI Models: A Financial and Technological Commitment}
\label{scaling_laws_in_africa}

In the context of AI development or research, scaling laws are empirical relationships that describe how the performance of AI models (particularly large language models) improves predictably as we increase their size, the amount of data on which they are trained, or the computing power used. 
In the modern AI era, these laws are crucial to understanding the limitations and potential of AI models, often prior to running experiments.\\
In their seminal work \cite{kaplan2020scaling}, \textit{OpenAI} argued that the training objective\footnote{In the context of machine learning, an objective refers to quantifying the goal the model is built to achieve during its training process using mathematical functions, optimization algorithms, and evaluation metrics to measure performance.} of a language model scales as a power-law with respect to model parameters, dataset size, and compute capabilities. 
Larger models or more data do not always result in better outcomes unless they are correctly balanced with the available computing resources. \\
Building upon \cite{kaplan2020scaling}, the \textit{Chinchilla} scaling laws, proposed by \textit{Google DeepMind} researchers, suggest that for a given compute budget, the optimal strategy for training large language models is to increase both the model size (number of parameters) and the training data size (number of tokens) at equal rates. 
Specifically, their findings show that a more balanced scaling of parameters and data leads to more compute-efficient and better-performing models.
This latter formulation of the scaling laws also emphasizes that beyond a certain point, increasing model size without proportionally increasing data does not yield significant performance gains.

\textbf{\textit{The Growing Cost of Cutting-Edge AI Innovation}}. The development and even the deployment of state-of-the-art AI models like \textit{OpenAI}'s GPT-4, \textit{DeepSeek}, or \textit{Meta}'s LLama underscores the critical role of substantial compute, extensive and quality datasets, or simply put: significant investment. 
These models have been trained on vast datasets using thousands of advanced GPUs, which therefore results in significant costs. 
Developing them reportedly involved spending between a few million and several hundreds of million US dollars. 
As an example, training LLama 3 405B \cite{grattafiori2024llama3herdmodels}, considering the 16K H100 GPUs used, can be roughly estimated at over US\$ 60 million, based on GPU rental costs of approximately US\$ 1,500 per GPU per month over a three-month period.\footnote{See \url{https://lambda.ai/pricing} for an estimate of the pricing. Renting H100 GPUs typically costs over US\$ 2.2 per GPU-hour, for reserved instances (September 2025).}\\
There have also been debates about the disclosed budget for the development of \textit{DeepSeek}-V3 \cite{deepseekV3}, estimated at roughly US\$ 5.6 million for the official training run for up to 2 months. 
Usually, these estimates primarily account for GPU rental/usage during the final training phases, omitting substantial investments in hardware acquisition, data center infrastructure maintenance, and data procurement. 
Moreover, the development process of new deep learning models has continuously involved extensive R\&D activities, including multiple experiments for hyperparameter tuning and ablation studies, which contribute significantly to the overall cost. 
Therefore, when considering the comprehensive infrastructure-related expenses, the actual cost for developing \textit{DeepSeek}-V3 likely far exceeds the publicly stated figures.\footnote{In their analysis, the authors pointed out that the estimated expenses solely cover the official training of \textit{DeepSeek}-V3, while omitting costs related to prior research and ablation experiments on architecture, algorithms, or data.}

Although limited domestic resources restrict the ability of local AI practitioners to develop frontier models, understanding scaling laws and their implications allows for strategic allocation of the available resources.

\textbf{\textit{Implications for Applied Research}}. Scaling laws may feel like a deterrent, but they don't preclude African nations from developing frontier AI. Rather, they underscore the unique strategic decisions required for full participation.
Training large models entirely from scratch like InkubaLM by \textit{LelapaAI}\footnote{https://lelapa.ai/} \cite{tonja2024inkubalm} may be financially or technically out of reach for many African AI specialists, but leveraging existing open-source models and adapting them to local languages and contexts offers a practical and impactful alternative as demonstrated by the \textit{Sunbird AI}\footnote{https://sunbird.ai/} team from Uganda \cite{akera2025sunflowernewapproachexpanding}. 

\textbf{\textit{Implications for Fundamental Research}}. AI breakthroughs have historically been fueled by the concurrent availability of relevant compute, datasets, and skilled professionals. 
Even modest investments in compute capacity can have outsized effects. 
A notable example is Nvidia's 2016 donation of one of the firsts DGX-1 supercomputer to \textit{OpenAI}, which drastically accelerated the firm's early breakthroughs. 
Similarly, African innovators don't necessarily need on-premise hyperscale infrastructure before starting building AI solutions for their communities.\\ 
Building on Africa's AI declaration \cite{africa_ai_declaration_2025}, each African country should further articulate their aspirations for engaging in both fundamental and applied AI research to ensure that investments are appropriately directed.\footnote{The declaration was adopted at the inaugural Global AI Summit on Africa in April 2025.}
For those with existing AI strategies, it involves clarifying their implementation roadmaps to facilitate collaboration on shared goals.

\subsection{Existing Archetypes in Advanced AI Ecosystems}
\label{global_examples}
Scaling AI infrastructure in Africa faces significant hurdles due to the region's fragmented and often unreliable energy landscape \cite{IEA2025}, gaps in mobile network coverage, and limited fixed broadband penetration \cite{ITU2025digitaldev}. 
Given the high power demands of AI development workloads, it is crucial for African countries to act carefully while concurrently addressing the pressing need for household electrification.
This delicate balance requires innovative solutions that ensure sustainable and equitable digital transformation.\\
In the following lines, we propose to explore some approaches and strategies to AI adoption by other regions, highlighting key insights that may be transferable to SSA.

\subsubsection{The European Union's Approach to Regional HPC Capabilities}
The European HPC Joint Undertaking (EuroHPC JU)\footnote{\url{https://www.eurohpc-ju.europa.eu/index_en}} represents an interesting model of regional cooperation. 
It strategically locates its compute centers near stable and renewable energy sources throughout the EU, prioritizes energy-efficient hardware, and builds infrastructure through cooperative continental efforts.
This partnership involves 36 European countries, for a budget of around EUR 8 billion from 2021 to 2027 \cite{eurohpc}.\\ 
\textit{JUPITER}, Europe's fastest (exascale) and most energy-efficient supercomputer, takes center stage in the EuroHPC JU network.\footnote{\url{www.fz-juelich.de/en/ias/jsc/jupiter}}
Powered by an astounding 24K of Nvidia's latest Grace Harper (GH200) chips, this powerhouse serves as a testament to the region's unwavering dedication to providing cutting-edge resources and support for local innovators.
This commitment reinforces Europe's position at the forefront of AI research, fostering innovation and progress in a variety of fields, contrasting the growing narrative about the continent's being mostly focusing on data/AI regulation.\\

\paragraph*{Key Takeaways: SSA could emulate this approach by fostering multi-country partnerships to fund, manage, and share AI infrastructure. 
Putting this into the perspective of the African declaration on AI, establishing a similar joint undertaking would require about 17\% of its planned AI fund (about 10 billion US\$).\\}

\subsubsection{The Asian Model: Robust Foundations and Government Support}
Asia's model for developing robust AI ecosystems, if there exists any, is characterized by strategic foundations and strong governmental support, exemplified by Taiwan, China, and Singapore.
Each country has adopted unique approaches to nurture its domestic AI landscape.

\textbf{\textit{Singapore}}. Ranked as the best in the world in human capital development,\footnote{2020 World Bank Human Capital Index, see \url{https://www.worldbank.org/en/publication/human-capital}.} Singapore has established initiatives such as the AI Apprenticeship Programme (AIAP) through AI Singapore,\footnote{\url{aisingapore.org}} which provides structured, paid training that seamlessly integrates academic learning with practical AI applications. 
This program illustrates how to effectively bridge the gap between education and industry demands. 
Furthermore, entities like SGInnovate\footnote{\url{www.sginnovate.com}} play a key role by co-investing in deeptech startups, creating a symbiotic relationship between government, academia, and the private sector. 
Singapore has continuously made efforts to improve its AI ecosystem through international collaborations, e.g., the partnership between the National University of Singapore and FPT Corporation for the establishment of a state-of-the-art AI lab.\footnote{\url{https://news.nus.edu.sg/fpt-nus-join-forces-in-driving-ai-innovation-fostering-talent-development/}}

\textbf{\textit{China}}. The Chinese approach to AI, as outlined in its 2017 \textit{New Generation Artificial Intelligence Development Plan}, emphasizes a strategic, state-driven vision to become a global leader in AI by 2030.
Its overall strategy is built on several foundational principles, including open-source, technology-led growth, massive R\&D and sectoral integration.
The plan prioritizes investments in core technologies such as machine learning (ML), natural language processing (NLP), and robotics, while fostering collaboration between academia, industry, and government \cite{webster2017chinaNGAIDP}. 
It also focuses on integrating AI into key sectors such as healthcare, transportation, defense, and manufacturing to drive economic growth and social progress. 
Another key aspect of the country's AI strategy is the establishment of AI research institutes supported by substantial public and private funding \cite{webster2017chinaNGAIDP}.\\
Foundational efforts, especially in talent pipelines, have significantly contributed to the growth of local firms like \textit{Alibaba}, \textit{ByteDance}, \textit{Baidu}, to name a few. 
The benefits were also noticed in \textit{DeepSeek}'s hiring process, which the firm argued was primarily targeting fresh graduates from top Chinese universities or PhD candidates nearing completion \cite{deepseek2024jordan}.\\
In recent years, China's strategy highlights a strong focus on attracting highly skilled individuals, in a global situation where many other countries, i.e., the US or Canada are tightening policies on immigration.
In essence, China's massive educational investment has produced vast numbers of graduates (especially in STEM), reaching about 39 million undergraduates in 2024.

\textbf{\textit{Taiwan}}. Taiwan adopts a strategic focus on specialization within its AI ecosystem.
Taiwan's AI Taiwan Action Plan (2018-2021) aimed to position the island as a global leader in AI by leveraging its strengths in semiconductors, fostering innovation, and integrating AI into industries. 
The plan emphasized actions like talent development (training researchers and producing 10K+ technicians annually), semiconductor leadership by expanding the country's global chip industry, and building an AI innovation hub in order to attract tech giants like Microsoft and Google. 
Using its strong semiconductor industry, Taiwan has established itself as a leader in AI hardware production. 
In particular, it is worth noting that the world increasingly relies on Taiwan Semiconductor Manufacturing Company (TSMC),\footnote{TSMC is a leading semiconductor manufacturer which pioneered the pure-play foundry business model, manufacturing chips for other firms rather than designing its own.} for advanced chip production serving major tech companies like Nvidia.
However, while this is a key advantage for the country, it is also creating significant dependence and posing a threat given the growing global demand for advanced AI-specific hardware and the global geopolitical situation.\\

\paragraph*{Key Takeaways: Together, these approaches from Asian countries highlight the importance of government involvement in setting clear policies, investing in education, and promoting public-private partnerships.\\}

\subsubsection{The Northern America Way: Strong Innovation, Fundamental and Applied Research}
In both the United States (US) and Canada, robust ecosystems for AI development have been cultivated through continuous government support, industry collaboration, and academia-led research.

\textbf{\textit{The United States}}. The country has long been a leader in AI innovation, driven by a combination of private sector dynamism and public investment. 
Key players include tech giants (Meta, Google, Microsoft, Nvidia, Amazon, Tesla, etc.) that invest heavily in AI R\&D and support ecosystem growth through their contributions to open research. 
On the one hand, government agencies such as the National Science Foundation (NSF) and Defense Advanced Research Projects Agency (DARPA) provide substantial grants and support for AI research, fostering cutting-edge advancements.
On the other hand, the US has continuously been home to startups that have significant impact on how humanity engages with advanced technologies, hosting firms like \textit{Hugging Face}, \textit{Scale AI} (Owned by \textit{Meta} at about 49\%), \textit{OpenAI}, \textit{Anthropic}, \textit{Skild AI}, \textit{Perplexity} and many others.\\
As reported by OECD's 2022 data on new ICT short-cycle tertiary graduates\footnote{\url{https://data-explorer.oecd.org}}, the US was way ahead in producing more graduates at each of the associate, bachelor's, master's, and PhD levels than any other country included in the data set \cite{maslej2025StanfordAIIndex2025}.
For instance, Carnegie Mellon University (CMU) secured the top spot among US academic institutions in 2023.\footnote{\url{https://nces.ed.gov/ipeds/use-the-data\#SearchExistingData}}
Distinguishing itself from others, CMU was one of the few universities that offered specialized programs in AI until relatively recently \cite{maslej2025StanfordAIIndex2025}. 
The university has also been instrumental in supporting the same dynamics in Africa by establishing a regional campus in Kigali, Rwanda.

\textbf{\textit{Canada}}. Over the years, Canada has emerged as a notable hub for AI innovation, supported by its forward-thinking policies and investments in education and research. 
The Canadian government launched the Pan-Canadian Artificial Intelligence Strategy, allocating funds to build AI capabilities across universities and research institutions. 
This strategy emphasizes collaboration between academia, industry, and government, creating an ecosystem that encourages innovation while prioritizing ethical considerations. 
Canada hosts AI research institutions such as the Montreal Institute for Learning Algorithms (MILA), a leading research institute renowned for its pioneering work in ML, particularly in deep learning. 
MILA is a hub for cutting-edge AI research and innovation, contributing significantly to global advancements in the field.
Alongside Toronto's Vector Institute and Alberta's Machine Intelligence Institute (Amii), it is one of the three national hubs of the Pan-Canadian AI Strategy.\footnote{https://ised-isde.canada.ca/site/ai-strategy/en}\\
\textit{Cohere} is another key player in Canada's AI landscape, acting as a national champion that attracts significant government and private investment.
The firm has raised approximately \$1.54 billion in 7 funding rounds since its founding in 2019, making it one of Canada's best-funded AI startups.\footnote{\url{https://www.canada.ca/en/innovation-science-economic-development/news/2025/03/government-of-canada-finalizes-investment-to-support-canadian-born-ai-leader-cohere.html}}\\

\paragraph*{Key Takeaways: The US and Canada exemplify the importance of public-private partnerships and investment in AI-focused education as cornerstones for the development of sustainable AI ecosystems.
Their strategies provide valuable insight into the balance of innovation and social responsibility, making them models worth emulating globally.\\}

\section{Africa's Path Forward}
\label{africas_path_forward}

\subsubsection{Accelerating and Adapting Capacity Building Efforts}
\label{accelerating_capacity_building}
Empowering Africans with advanced AI skills is essential to unlock the continent's potential and foster a vibrant ecosystem of talented professionals who can contribute significantly to both local and global AI innovation. 
To this end, efforts must be redoubled to better equip early-career researchers with the necessary expertise to transform the youthful population into skilled contributors to AI and related fields.
Achieving this goal requires not only providing access to education and training, but also creating an environment that encourages and retains local talents. 
Incentives such as valuing research through grants allocations, decent salaries, and opportunities for career advancement can go a long way to attracting and retaining top talent in the region.

Many academic institutions across Africa are contributing to cultivating technical talent.
Community-driven efforts like the Masakhane project\footnote{\url{www.masakhane.io}} demonstrate the impact of grassroots collaboration in developing AI tools for African languages.

\textbf{\textit{Academic Institutions}}

The Mohammed VI Polytechnic University (UM6P) in Morocco, which has established an International Center for AI to foster Moroccan expertise in AI and Data Sciences. \\
CMU-Africa, a partnership between Carnegie Mellon University (CMU) and the government of Rwanda, offers a master's program in engineering artificial intelligence that equips learners with advanced skills that cover key sectors such as transportation, energy and healthcare. \\
The University of Pretoria (UP) in South Africa hosts the data science for social impact (DSFSI) group,\footnote{\url{https://www.dsfsi.co.za/}} a research group that has been a very active contributor to Africa's AI advancement, particularly in African language NLP and ML research.\\
The African Institute for Mathematical Sciences (AIMS), with centers in South Africa, Rwanda, Ghana, Cameroon, and Senegal, offers advanced training in AI and machine learning, including programs like the AI for Science Master's in partnership with Google DeepMind.

\textbf{\textit{Grassroots AI Communities}}
\label{africa_ai_talents}

Besides efforts by academic institutions, \textit{Deep Learning Indaba (DLI)}\footnote{Deep Learning Indaba is the largest AI-focused conference on the African continent.} serves as a continental effort for strengthening AI capacity by organizing regular gatherings across Africa, promoting collaboration and knowledge sharing among local researchers, AI practitioners, and especially students.\\
Local AI communities like \textit{Galsen AI} in Senegal, \textit{AI Kenya},\footnote{\url{https://kenya.ai}} \textit{Mbaza NLP} in Rwanda, \textit{Data Science Nigeria},\footnote{\url{https://datasciencenigeria.org}}, and numerous others are working diligently to promote widespread access to AI throughout the region. 
These communities primarily consist of young developers, researchers, and students.\\
Examination of recent data from DLI \cite{dli2023report, dli2024report} reveals that Africa's most active AI practitioners are predominantly students at both undergraduate and postgraduate levels. 
This contrasts with traditional African research habits, which have typically relied on PhD students and senior researchers to drive academic endeavor. 
The shift reflected in DLI reports highlights a growing trend toward empowering early-career researchers\footnote{Early career researchers are scientists or scholars in the initial stages of their research careers who may be pursuing advanced degrees such as Master's or PhDs; recent graduates who have started working on research projects independently for the first time.} (ECRs).
Enhancing opportunities for young researchers to rise through the ranks and take on prominent roles can further strengthen Africa's burgeoning AI ecosystem.

In addition to AI-specific initiatives, valuable lessons can be drawn from the continent's growing robotics ecosystem, which is evolving despite similar constraints. 
Programs like the African Robotics Network (AFRON) and its successor, AfRob, alongside regional robotics competitions and training efforts by institutions such as Ashesi University, Fundi Bots in Uganda, and the Rwanda National Robotics Program, have demonstrated the power of community-driven innovation in the face of infrastructure scarcity \cite{vernon2025robotics}. 
These robotics initiatives emphasize the importance of hands-on education, low-cost but efficient hardware, and regional collaboration, all of which are equally critical to fostering local talent and encouraging homegrown solutions. 
Integrating robotics, AI, and embedded systems training across African educational systems, from basic to advanced levels, could help build a more robust pipeline of practitioners equipped to navigate and close the digital divide.

\subsubsection{Breaking the Cycles of Short-term Efforts}
\label{braeking_cycles_of_short_term_efforts}

\begin{figure*}[!ht]
    \begin{subfigure}{0.5\textwidth}
\pgfplotstableread[row sep=\\,col sep=&]{
    company & private & public & hpc \\
    Meta (US)              & 16000  & 5400  &   0  \\
    Tesla (US)             & 16000  &  0    &   0 \\
    Leonardo (EU)          & 13800  &  0    &  0  \\
    DeepSeek (CHIN)         & 10000  &  0    &  0  \\
    Lambda (US)            & 0      & 10000 &  0\\
    XTX Markets (UK)       & 10000  & 0     &  0\\
    Perlmutter (US)         & 0      & 0     &  6160\\
    Stability.ai (UK)      & 0      & 5410  &  0\\
    NVIDIA Selene (US)     & 4320   & 0     &  0\\
    JUWELS (DE)            & 0      & 0     &  3740\\
    Polaris (US)           & 0      & 0     &  2400\\
    Hugging Face (US)      & 0      & 1050  &  0\\
    Aleph Alpha (DE)       & 532    & 512   &  0\\
    Contextual.ai (US)     & 0      & 1020  &  0\\
    MeluXina (EU)          & 0      & 0     &  800\\
    NVIDIA Cam-1 (UK)      & 640    & 0     &  0\\
    Narval (CA)            & 0      & 0     &  636\\
    Hessian.ai (DE)        & 0      & 0     &  632\\
    Karolina (EU)          & 0      & 0     &  576\\
    Tursa (UK)             & 0      & 0     &  456\\
    Jean Zay (FR)          & 0      & 0     &  416\\
    Recursion (US)         & 340    & 0     &  0\\
}\acloud

\resizebox{1.\linewidth}{!}{
\begin{tikzpicture}
\begin{axis}[
    ybar,
    width=1.6\linewidth,
    bar width=5pt,
    height=8cm,
    ylabel={A100 Count},
    ymin=0,
    ymax=20000,
    symbolic x coords={
        Meta (US),  
        Tesla (US), 
        Leonardo (EU), 
        DeepSeek (CHIN),
        Lambda (US),
        XTX Markets (UK),
        Perlmutter (US),
        Stability.ai (UK),
        NVIDIA Selene (US),
        JUWELS (DE),
        Polaris (US),
        Hugging Face (US),
        Aleph Alpha (DE),
        Contextual.ai (US),
        MeluXina (EU),
        NVIDIA Cam-1 (UK),
        Narval (CA),
        Hessian.ai (DE),
        Karolina (EU),
        Tursa (UK),
        Jean Zay (FR),
        Recursion (US)
    },
    xtick=data,
    xtick align=center,
    xticklabel style={anchor=east, rotate=90, font=\footnotesize},
    axis lines=left,
    enlarge x limits=0.04,
    nodes near coords,
    nodes near coords style={
        font=\footnotesize,
        rotate=90,
        anchor=west
    },
    every node near coord/.append style={
        /pgfplots/number format/.cd,
    },
]

\addplot+[draw=none, forget plot] coordinates {
    (Meta (US), -1)
    (Tesla (US), -1)
    (Leonardo (EU), -1)
    (DeepSeek (CHIN), -1)
    (Lambda (US), -1)
    (XTX Markets (UK), -1)
    (Perlmutter (US), -1)
    (Stability.ai (UK), -1)
    (NVIDIA Selene (US), -1)
    (JUWELS (DE), -1)
    (Polaris (US), -1)
    (Hugging Face (US), -1)
    (Aleph Alpha (DE), -1)
    (Contextual.ai (US), -1)
    (MeluXina (EU), -1)
    (NVIDIA Cam-1 (UK), -1)
    (Narval (CA), -1)
    (Hessian.ai (DE), -1)
    (Karolina (EU), -1)
    (Tursa (UK), -1)
    (Jean Zay (FR), -1)
    (Recursion (US), -1)
};
    
\addplot+[ybar, bar shift=-0.1cm] table[x=company,y=private,
restrict expr to domain={\thisrow{private}}{1:20000}
]{\acloud};

\addplot+[ybar, bar shift=0.12cm] table[x=company,y=public,
restrict expr to domain={\thisrow{public}}{1:20000}
]{\acloud};

\addplot+[ybar, bar shift=0.01cm] table[x=company,y=hpc,
restrict expr to domain={\thisrow{hpc}}{1:20000}
]{\acloud};

\end{axis}
\end{tikzpicture}
}
        \caption{A100 GPU counts}
        \label{fig:a100_counts}
    \end{subfigure}  
    \begin{subfigure}{0.5\textwidth}
\pgfplotstableread[row sep=\\,col sep=&]{
    company & private & public & hpc \\
    Meta (US)              & 350000 & 0       &  0   \\
    XAI (US)               & 100000 & 0      & 0   \\
    Tesla (US)             & 35000  & 0     &  0  \\
    Lambda (US)            & 0      & 30000 &  0\\
    Google A3 (US)         & 0      & 26000 &  0\\
    Oracle Cloud (US)      & 0      & 16000 &  0\\
    Poolside (US)          & 0      & 10000 &  0\\
    Magic (US)             & 0      & 8000  &  0\\
    Yotta (IN)             & 0      & 10000 &  0\\
    Andromeda (US)         & 3630   &  0    &  0\\
    Scaleway (FR)          & 3000   &  0    &  0\\
    Hugging Face (US)      & 0      & 768   &  0\\
    DeepL (DE)             & 544    & 0     &  0\\
    Recursion (US)         & 504    & 0     &  0\\
    Photoroom (FR)         & 256    & 0     &  0\\
    Instadeep (UK)         & 224    & 0     &  0\\
    CSCS Alps (CH)         & 0     & 0     &  10800\\
    MareNostrum 5 (ES)     & 0     & 0     &  4480\\
    JUPITER (DE)           & 0     & 0     &  2000\\
    Gefion (DK)            & 0     & 0     &  1530\\
    Jean Zay (FR)          & 0     & 0     &  1460\\
    EPFL (CH)              & 0     & 0     &  336\\
    Princeton (US)         & 0     & 0     &  300\\
    Hessian.ai (DE)        & 0     & 0     &  280\\
    RWTH Aachen (DE)       & 0     & 0     &  208\\
    Ruhr Bochum (DE)       & 0     & 0     &  56\\
    Chan-Zuckerberg (US)   & 0     & 0     &  48 \\
}\hcloud

\resizebox{1.\linewidth}{!}{
\begin{tikzpicture}
\begin{axis}[
    ybar,
    width=1.6\linewidth,
    bar width=5pt,
    height=8cm,
    ylabel={H100 Count},
    ymin=0,
    ymax=400000,
    symbolic x coords={
        Meta (US), 
        XAI (US), 
        Tesla (US), 
        Lambda (US), 
        Google A3 (US), 
        Oracle Cloud (US), 
        CSCS Alps (CH), 
        Poolside (US),
        Magic (US), 
        MareNostrum 5 (ES),
        Yotta (IN), 
        Andromeda (US), 
        Scaleway (FR),
        JUPITER (DE),
        Gefion (DK),
        Jean Zay (FR),
        Hugging Face (US),
        DeepL (DE), 
        Recursion (US), 
        EPFL (CH),
        Princeton (US),
        Hessian.ai (DE),
        Photoroom (FR), 
        Instadeep (UK), 
        RWTH Aachen (DE),
         Ruhr Bochum (DE),
        Chan-Zuckerberg (US),
    },
    xtick=data,
    xticklabel style={anchor=east, rotate=90, font=\footnotesize},
    axis lines=left,
    enlarge x limits=0.03,
    nodes near coords,
    nodes near coords style={rotate=90, anchor=west, font=\footnotesize},
]

\addplot+[draw=none, forget plot] coordinates {
    (Meta (US),  -1)
    (XAI (US),  -1)
    (Tesla (US),  -1)
    (Lambda (US),  -1)
    (Google A3 (US),  -1)
    (Oracle Cloud (US),  -1)
    (CSCS Alps (CH),  -1)
    (Poolside (US), -1)
    (Magic (US),  -1)
    (MareNostrum 5 (ES), -1)
    (Yotta (IN),  -1)
    (Andromeda (US),  -1)
    (Scaleway (FR), -1)
    (JUPITER (DE), -1)
    (Gefion (DK), -1)
    (Jean Zay (FR), -1)
    (Hugging Face (US), -1)
    (DeepL (DE),  -1)
    (Recursion (US),  -1)
    (EPFL (CH), -1)
    (Princeton (US), -1)
    (Hessian.ai (DE), -1)
    (Photoroom (FR),  -1)
    (Instadeep (UK),  -1)
    (RWTH Aachen (DE), -1)
    (Ruhr Bochum (DE), -1)
    (Chan-Zuckerberg (US), -1)
};

\addplot+[ybar, bar shift=-0.04cm] table[x=company,y=private,
restrict expr to domain={\thisrow{private}}{1:400000}
]{\hcloud};

\addplot table[x=company,y=public,
restrict expr to domain={\thisrow{public}}{1:35000}
]{\hcloud};

\addplot+[ybar, bar shift=0.01cm] table[x=company,y=hpc,
restrict expr to domain={\thisrow{hpc}}{1:15000}
]{\hcloud};

\legend{Private Cloud, Public Cloud, National HPC}
\end{axis}
\end{tikzpicture}
}
        \caption{H100 GPU counts}
        \label{fig:h100_counts}
    \end{subfigure}
    \caption{Nvidia A100 and H100 GPUs across public, private, and national HPC infrastructure (redrawn from the state of AI Report - compute index 2024).}
    \label{fig:gpu_counts}
    \centering
\end{figure*}
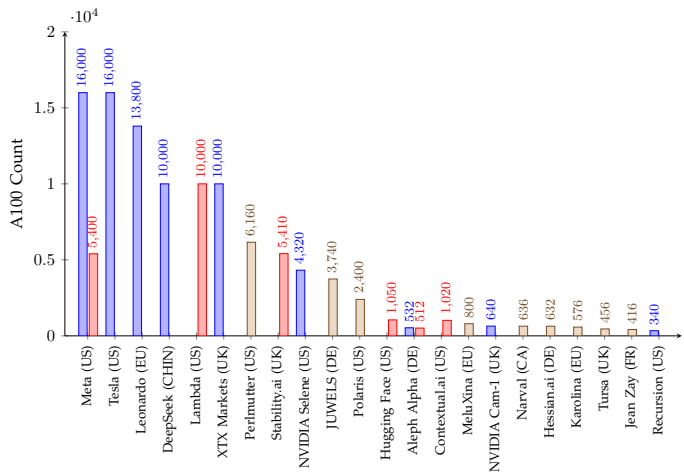
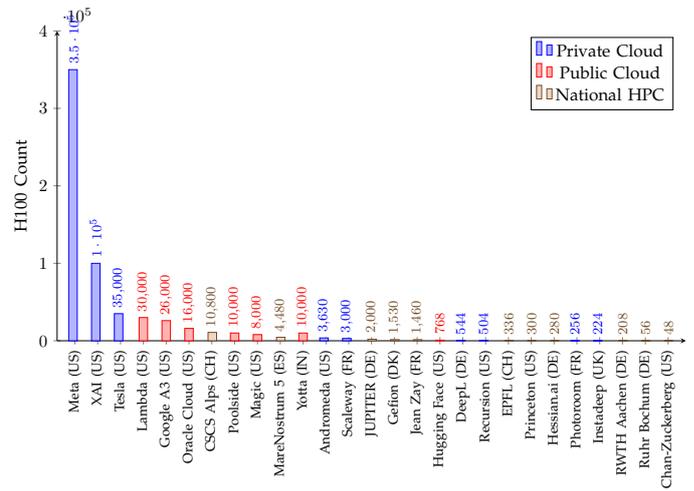

Despite the deep potential and ambitious experiments, the uneven capacity and the lack of strategic scaling frameworks hinder the transition from pilot to widespread adoption \cite{aiafrica2023state,gsma2024ai}. 
Looking at some of the factors influencing post-hackathon project continuation in the region, Ratsoga and Primus \cite{ratsoga2023hackathons} argue that discontinuity is largely due to insufficient funding, skills gaps, weak infrastructure, limited policy support, and poor integration of projects into existing systems. 
This suggests existing local creativity, but also highlights a persistent gap between ideation and operational maturity in Sub-Saharan Africa's digital ecosystem. 
These short-term innovation cycles, often driven by visibility goals or donor incentives, rarely transition to scalable, production-ready systems.\\
AI-focused entrepreneurship in Africa is growing but remains structurally constrained. Although more than 200 deep-tech startups have emerged across the continent (79\% of which were founded post 2014), the majority operate in early stages with limited access to funding, infrastructure, or specialized support. 
AI and machine learning represent the most active tech cluster within deep-tech, with applications spanning agriculture, healthtech, fintech, and logistics. 
The gap between university research and market-ready talent also hampers scalability \cite{intelcorp2022deeptech}.\\
Funding to AI startups in Africa is increasingly becoming more concentrated, as countries like Kenya, Tunisia, Egypt, South Africa, and Nigeria reportedly accounted for over 90\% of total investments in AI on the continent by H1 2025.\footnote{https://blog.startuplist.africa/articles/ai-revolution-in-africa-2025}
Although this trend highlights the increasing interest from global investors in the region's burgeoning AI landscape and offers promising opportunities for early-stage AI companies, there is still a need to address the cyclical nature of short-term projects that creates a sense of recurring restart rather than sustained growth.

To foster long-lasting development in AI, it is essential to focus on establishing consistent and reliable funding mechanisms for startups across the continent. 
This could involve creating an environment for venture capital (VC) funds dedicated exclusively to AI investments, as also suggested by Africa's declaration on AI with a continental AI fund \cite{africa_ai_declaration_2025}, and offering government grants for promising projects as well as implementing tax incentives for companies investing in AI R\&D.\\
Several African countries have already made strides by establishing startup acts or dedicated legal frameworks aimed at fostering the activities of domestic startups and SMEs.\footnote{Countries with such laws include Tunisia, Senegal, Nigeria, Côte d'Ivoire, Ethiopia, and Ghana.}
However, the impact on the ground remains limited. 
Bridging this gap requires addressing potential obstacles such as bureaucratic red tape, lack of enforcement resources, and insufficient coordination between government agencies and stakeholders in the private sector.

\subsubsection{A Sustainable Model for AI Compute in SSA}
\label{african_ai_compute_model}

\textbf{\textit{Building or Renting Computing Power}}.
\label{strategic_analysis}
Given global trends in AI supercomputers \cite{pilz2025trendsaisupercomputers,stateofaireport2025}, rented AI compute via global hyperscalers offers Africa a more realistic solution. 
However, the stark policy questions are the following: can Africa afford to delay local HPC development for years, risking falling behind in the global race to AI adoption? or must the continent prioritize immediate engagement with compute providers to harness AI's transformative potential, even at the cost of short-term compromises? 

The reality, as illustrated by Figure \ref{fig:gpu_counts}, is that very few national HPC installations worldwide match the scale, speed, and reliability of private/public cloud platforms, making rented AI compute a necessary bridge to avoid missing the AI revolution.

\textbf{Case Study}. An AI research team of four requires access to an 8-GPU server. 
The team usage pattern is mixed:
    \begin{itemize}
        \item Four GPUs are reserved for continuous weekly/monthly long runs.
        \item Each of the four researchers has a dedicated GPU for daily experiments and ablations that can potentially be scaled to the full server.
        \item The team works on various language technology projects with shared resources.
    \end{itemize}

The following analysis aims to determine the most cost-effective approach between purchasing (on-premise) and renting from cloud providers, considering both short-term (up to 6 months) and medium-term (1-3 years) projects.
It focuses on the efficiency and direct cost-effectiveness of each solution, and intentionally overlooks policy-related factors such as strategic/operational control, data and security considerations, reliability (latency and availability of stable broadband internet), or vendor lock-in to name a few.\\
Prices were estimated based on announcements by Nvidia and consultation of vendor websites. 
Additional details can be found in Appendix \ref{app:business_case}.

\textbf{Cost-Benefit Analysis}.
\label{cost_benefit_analysis}
Based on the described case, we argue that the optimal strategy for AI compute mainly depends on project duration and utilization rates.

\paragraph{Buying}
Purchasing an on-premise system involves significant investment and hidden costs. 
The initial hardware purchase alone for a pre-configured 8xA100 server might range from US\$ 150K to US\$ 250K, while a high-demand 8xH100 system can easily cost between US\$ 300K and US\$ 500K. 
Beyond the sticker price, annual operating costs quickly accumulate. 
Power and cooling can add US\$ 10K to US\$ 30K per year, depending on energy rates and system load. 
Enterprise support contracts and maintenance might add another US\$ 1K to US\$ 50K annually. 
Crucially, the cost of dedicated in-house DevOps or IT professionals to manage cluster infrastructure can range from US\$ 120K to US\$ 200K per year. \\
The estimated total cost of ownership (TCO) over three years for an on-premise system can easily exceed US\$ 500K for an A100 setup and approach US\$ 800K-1M for an H100 setup, assuming consistent high utilization.\\

\paragraph{Renting}
Rented cloud computing resources convert a large capital expense into a manageable operational expense, offering scalability and immediate access to hardware. 
On-demand hourly rates for an 8xA100 instance generally fall between \$7 and \$40 per hour, while 8xH100 instances range from approximately \$17 to \$98 per hour, depending on the provider and region. 
Utilizing these instances consistently to match the team's needs (daily experiments plus long runs, estimated at $\sim$70\% utilization) translates to an annual cost of roughly \$43K to \$245K for A100s, or \$103K to \$600K for H100s.\footnote{We assume $24*365$ hours in a year to which we apply 70\% as a realistic estimate for a busy team runing daily experiments and jobs.}
Note that for stable, long-running projects, major cloud providers offer reserved instances with 1- or 3-year commitments, which can reduce on-demand costs by 30-60\%. 
Furthermore, highly discounted spot instances (often less than \$10/hour for 8xA100) are perfect for fault-tolerant experimental tasks.\\

\paragraph{Conclusion}
For an AI team requiring an 8-GPU server, renting is generally a more cost-effective solution compared to purchasing on-premise hardware. 
The high initial capital expenditure (CapEx) for physical servers, combined with ongoing operational expenses (OpEx) for power, cooling, maintenance, and specialized IT staff, often makes on-premise ownership unfeasible for dynamic research environments. 
In the contrary, cloud platforms mitigate these issues by offering flexibility, immediate access to the latest technology, and variable pricing models that align better with fluctuating research demands.\\
This analysis based on a small-scale AI team can be extrapolated to larger teams, startup ecosystems, or country-wide HPC requirements by university researchers similar to the US National Science Foundation (NSF)-funded Advanced Cyberinfrastructure Coordination Ecosystem: Services \& Support (ACCESS) program.\footnote{\url{https://access-ci.org/}}
Although we do not share such estimates, the overall discussion about buying versus renting shows that even for a focused AI team, significant investments can be required when choosing either routes.

\textbf{A context-aware strategy}. Unless a rigorous business plan demonstrates a clear return on investment (ROI) and a path to sustainability, the purchase of on-premise AI compute is hardly justifiable at both small and large scales. 
In developing contexts where foundational needs such as universal healthcare access and food security remain critical challenges, such capital expenditure (capEx) decisions must prioritize tangible socioeconomic returns over long-term and policy-related goals like achieving digital sovereignty.
Moreover, these investments are contingent on the presence of a critical mass of skilled workforce needed to manage and utilize such advanced infrastructure effectively, which is currently a scarce resource in most countries in the SSA region.
All these conditions make GPU cloud renting a far more rational and responsible choice in the current context, at least in the medium term.

\textbf{Our recommendation}. A \textit{pragmatic hybrid} African AI compute model.

To address immediate computational needs (up to 6 months) and bridge the gap in local HPC infrastructure, African nations may want to prioritize renting. 
This approach allows developers and researchers to access cutting-edge AI compute resources at scale while minimizing upfront costs, and leveraging existing global infrastructure to avoid delays in adoption. 

In the medium term (1-3 years), countries can transition to building modular, scalable AI factories that integrate renewable energy sources to ensure sustainability. 
Projects like \textit{Cassava AI}'s network of AI factories \cite{Nvidia2024cassava} and Uganda's \textit{Aeonian Project} demonstrate Africa's capacity to deploy such HPC facilities through public-private partnerships. 
Nevertheless, it should be noted that even though on-premise hardware might seem viable for continuous runs, the applicable discounts on reserved instances for commitments in the mid-term range make renting even more cost-effective in the medium term.
Thus, with carefully planned agreements on cloud-based infrastructure, building domestic HPC facilities may not come at the expense of missing on the ongoing AI race or mostly consuming foreign solutions.

In the long run, it is in the best interest for African nations to aim for joint undertaking approaches, e.g. \textit{EuroHPC JU}, in a unified strategy that ensures compute resources are shared across nations, at least those within the same regional economic community (REC). 
A pooled multi-country approach has higher chances to enable African governments and RECs to negotiate more favorable terms with providers. However, this long-term vision requires trust-building and interoperability frameworks to enable seamless collaboration, ensuring that Africa's overall AI strategy remains both autonomous and grounded in local policies.

\section{Limitations and Future Work}
\label{limitations}
This paper broadly analyzed the infrastructure, and policy barriers contributing to the perceived digital divide in African AI development. 
It is fair to acknowledge some limitations.
\begin{itemize}
    \item We recognize that our reliance on public data may not reflect the full scope, as it does not directly include proprietary infrastructure or unpublished initiatives from the private sector. 
    In fact, many of the deployed AI-ready compute resources either by African governments or local private firms remain under-documented or non verifiable.
    \item Planned installations or those in construction during the drafting of this paper were not included in the initial version of the Africa compute tracking tool presented in section \ref{act}.
    \item Our analysis is solely based on the Nvidia ecosystem, as it is (to date) widely considered the standard for AI research and development globally, primarily due to its dominant GPU hardware and proprietary CUDA\footnote{The Compute Unified Device Architecture (CUDA) is a parallel computing platform developed by Nvidia for its GPUs.} software platform \cite{bicaj2025NvidiaDominance}. 
    Notable alternatives like Open Computing Language (OpenCL), AMD's RoCm platform and its Heterogeneous-Compute Interface for Portability (HIP), or Intel's oneAPI were not considered in our analysis.
    \item This work focuses predominantly on the compute infrastructure, model scalability, and workforce availability as enablers of AI development. Thus, it puts less emphasis on the ethical, linguistic and socio-cultural dimensions, which we recognize are equally crucial for creating inclusive AI ecosystems in Africa.
\end{itemize}

\vspace{-3mm}
Moreover, we do not cover other types of AI accelerator hardware beyond GPUs, which we selected according to the ongoing trend for AI adoption in SSA. 
We believe this intentional omission presents a significant avenue for further exploration and research around AI compute capability in the region. 
Hardware such as Tensor Processing Units (TPUs), Neural Processing Units (NPUs), Field-Programmable Gate Arrays (FPGAs), Language Processing Units (LPUs), and Application-Specific Integrated Circuits (ASICs) are gaining tranction in the advancement of AI technologies. 
And an extension of our work towards these areas could provide valuable insights into optimal hardware choices for specific AI applications within the region, ultimately enhancing AI adoption.

\section{Conclusion}
\label{conclusion}
To close the growing AI divide, Africa must address infrastructure, accessibility, and affordability in tandem. 
While strategic and policy frameworks are essential, they, alone, cannot bridge the gap without tangible, coordinated actions by all key stakeholders. 
The proposed taxonomy of barriers to AI adoption in sub-Saharan Africa and the \textit{Africa AI Compute Tracker (ACT)} are some of our key contributions to providing data-driven tools as a foundation for targeted investments in AI compute resources, energy systems, and domestic cloud platforms.
The success of these efforts hinges on overcoming systemic barriers (in governance, infrastructure, human capital, and accessibility), which have historically limited the scalability of nascent AI initiatives. 
Through genuine regional collaborations, Africa can align its AI ambitions with global standards while still prioritizing local needs. 
The urgency of this task cannot be overstated: without immediate, action-oriented strategies to accelerate the development of digital foundations, Africa risks being left behind in the global AI adoption movement.\\
Our proposed AI Compute Tracker is not just a monitoring tool, but a call to action—a step toward ensuring that Africa's AI future is both equitable and sustainable, but mostly led by local resources.

\bibliographystyle{IEEEtranN}
\bibliography{citations}

\cleardoublepage
\onecolumn
\raggedbottom
\label{appendices}
\section{Appendix}
\addcontentsline{toc}{section}{Appendix}

\subsection{Definitions adopted}

\textbf{\textit{Artificial intelligence (AI)}} mainly refers to machine learning (ML) or deep learning-based techniques used to enhance data-driven processes and automate high-value tasks. In particular, we do not use the term ``AI'' as a proxy to generative AI.

A \textbf{\textit{graphics processing unit (GPU)}} is a specialized computer processor designed for parallel computation, and commonly used for rendering graphics and accelerating ML computations. 

\textbf{\textit{AI Compute}} refers specifically to computing capabilities or the computing power required to run advanced AI algorithms, usually in a distributed manner. In this context, the term ``cloud'' is employed as a proxy to ``public cloud''.

\textbf{\textit{Node}} is the term used for a single computer or server in a network that can perform tasks independently but can also communicate with other units for coordinated functioning in a distributed system.

\textbf{\textit{Cluster}} denotes a group of connected and interoperable nodes working together to achieve common goals such as enhancing computational power, and facilitating large-scale processing.

\textbf{\textit{AI Talent}} is a group of skilled individuals with demonstrated—not potential—expertise in AI-related topics and the ability to design, develop, and implement AI solutions; highly sought after for their innovative problem-solving abilities in diverse domains.

\textbf{\textit{AI model training}} refers to the process of teaching machine learning models from (large) datasets to automatically learn to solve tasks of interest.

\subsection{A Snapshot of Supercomputers in the TOP500 List}
\label{top_countries}


\begin{figure*}[ht]
\centering
\resizebox{.95\textwidth}{!}{

\begin{tikzpicture}
\begin{axis}[
    xbar,
    width=\textwidth,
    height=12cm,
    xlabel={Number of supercomputers},
    ytick=data,
    axis lines=left,
    yticklabel style={text width=4cm, align=center},
    symbolic y coords={
        Turkey, Bulgaria, Thailand, Czechia, Austria, United Arab Emirates, Spain, Finland, Ireland, Singapore,
        Australia, Switzerland, Norway, Russia, India, Saudi Arabia, Taiwan, Poland, Sweden, Canada, Brazil,
        Netherlands, South Korea, Italy, United Kingdom, France, Japan, Germany, China, United States
    },
    nodes near coords,
    xmin=0,
    bar width=5pt,
    enlarge y limits=0.02,
    title={}
]
\addplot coordinates {
    (2,Turkey)(2,Bulgaria)(2,Thailand)(3,Czechia)(3,Austria)(3,United Arab Emirates)
    (3,Spain)(3,Finland)(4,Ireland)(4,Singapore)(4,Australia)(5,Switzerland)(6,Norway)
    (6,Russia)(6,India)(7,Saudi Arabia)(7,Taiwan)(8,Poland)(8,Sweden)(9,Canada)(9,Brazil)
    (10,Netherlands)(13,South Korea)(14,Italy)(14,United Kingdom)(24,France)(34,Japan)
    (40,Germany)(63,China)(173,United States)
};

\end{axis}
\end{tikzpicture}
}
\caption{Distribution of supercomputers in the TOP500 list by country (as per November 2024 data)\protect\footnotemark.}
\label{fig:top500_distribution}
\end{figure*}
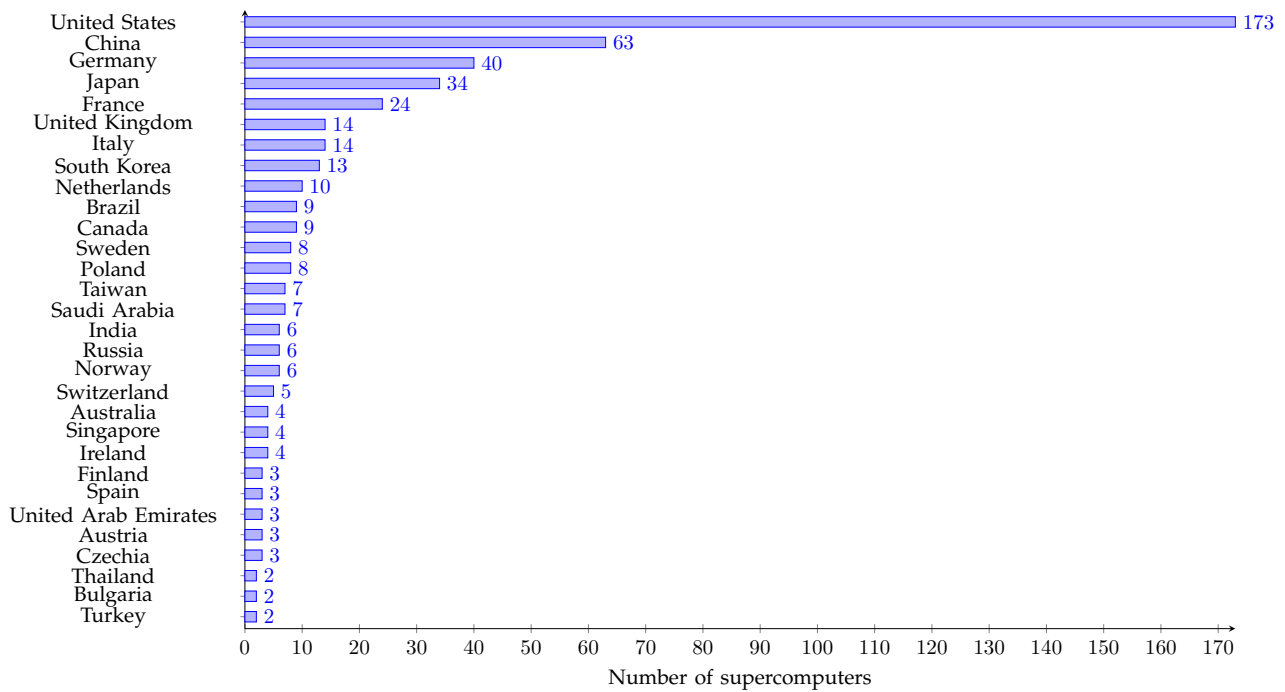
\footnotetext{\url{https://top500.org/lists/top500/list/2024/11/}}

\subsection{Detailed Business Case: Buying vs. Renting AI Compute}
\label{app:business_case}
This analysis is built around the 2025 1-month long Kinyarwanda automatic speech recognition (ASR) hackathon\footnote{\url{https://digital-umuganda.github.io/kasr_hackathon/}} hosted by the Rwandan company \textit{Digital Umuganda} in collaboration with the \textit{Gates Foundation}.
It is an example of a short-term, high-intensity project, which we can be scaled up for medium-term language technology development.

\textbf{Overview}. The competition provided a large-scale, multilingual dataset for Kinyarwanda, spanning Health, Government, Financial Services, Education, and Agriculture to support robust ASR model development in both conversational and formal contexts. 
The dataset is divided into three tracks: Track A (540 hours of transcribed speech), Track B (1180 hours of transcribed speech), and Track C (1180 hours of transcribed + 1170 hours of unlabeled audio), catering to varying levels of expertise and resource availability.\\
Submissions to this competition were scored using Word Error Rate (WER) and Character Error Rate (CER), combined into an Overall Score ($0.4*WER + 0.6*CER$). 
Higher scores indicate better performance, with emphasis on both word-level and character-level accuracy.

\textbf{Constraints}
\begin{itemize}
    \item Track A: Required use of only the provided transcribed data; no external corpora or cross-lingual pre-training allowed.
    \item Track B: Similar restrictions as Track A but allows up to \textbf{300 GPU hours}; external data is prohibited.
    \item Track C: Encourages semi/self-supervised learning with unlabeled data (up to \textbf{400 GPU hours}), permits external open-source datasets (e.g., Common Voice), and mandates model size disclosure (parameters, inference real-time factor/RTF). Private or paid datasets are banned.
\end{itemize}

The below scoping of experiments and compute requirements is based on reports from wining teams namely \textit{ASRwanda}\footnote{\url{https://www.kaggle.com/competitions/kinyarwanda-automatic-speech-recognition-track-a/writeups/asrwanda-1st-place-solution-for-kinyarwanda-asr-tr}} (Track A) \textit{Sunbird AI}\footnote{\url{https://www.kaggle.com/competitions/kinyarwanda-automatic-speech-recognition-track-b/writeups/sunbird-1st-place-solution-for-kinyarwanda-automat}} (Track B \& C).
The reported solutions did not include in-depth details on failed trials or ablation rounds, which are relevant for accurate estimates.

\subsubsection{Project Scenarios \& Compute Needs}
The development of African language technologies involves diverse workloads, from data processing to fine-tuning and training models from scratch. 

We assume the need for high-end hardware (Nvidia A100s/H100s) and a typical 8-GPU system for large-scale training or fine-tuning.
These compute resources are to be shared across a team of 4 AI researchers/developers running daily experiments and longer (weekly/monthly) jobs using specific sets of GPUs in an agile fashion.\\
Storage requirements are ignored, although it is fair to acknowledge their relevance in the design of the development environment for AI projects.

\textbf{Short-Term Project (Up to 6 months)}.
Experiments reported by wining solutions lasted 30-96 hours.
Since the goal is rapid prototyping and benchmarking, these experiments require flexible, on-demand access to multi-GPU instances as much as possible.
These projects usually have bursty, unpredictable usage patterns, and focus on speed and flexibility.

\textbf{Medium-Term Project (1-3 years)}.
We define a medium-term project in the same line as building a comprehensive suite of Kinyarwanda speech/text models. 
This involves extensive training from scratch and fine-tuning, potentially with experiments up to 96 hours or more.
As opposed to short-term ones, these projects are more consistent with predictable workloads, and have high potential for 24/7 GPU utilization. 

\subsubsection{Cost Analysis}
Details about pricing for on-demand cloud GPUs are shared in table \ref{tab:on_demand_gpu_pricing} and a breakdown of the cost analysis is presented in table \ref{tab:app_cost_benefit_analysis}.

\textbf{On-Demand GPU Cloud Offerings}

\textbf{Pricing}.
Specialist GPU cloud platforms generally offer lower headline prices for the GPU hardware itself, which is crucial for controlling operational expenses (OpEx) for short-term projects or early-stage development with unpredictable workloads.
Regarding total cost of ownership (TCO), some specialist platforms often price the GPU separately from other resources (CPU, RAM, and storage), therefore adding to the total cost, while hyperscalers often bundle these elements into a single instance price.

\textbf{Reliability}. 
Marketplace providers like \textit{Vast.ai} and platforms like \textit{Thunder Compute} offer some of the lowest prices but are often interruptible (spot instances), meaning jobs might be stopped with short notice. 
Stable, dedicated instances from providers like {Lambda Labs} or hyperscalers generally cost more but offer production reliability.

\textbf{Networking and Egress}.
Hyperscalers typically charge significant data transfer (egress) fees, which can add 20-40\% to a monthly bill when moving large datasets or model weights. 
In comparison, some specialist providers offer transparent or waived egress fees.

Although prices are competitive, securing large quotas of H100s, for instance, can sometimes involve waitlists or approval processes, particularly with major cloud providers and especially when requests come from Africa.

\begin{table*}[!ht]
\centering
\caption{Comparison of various on-demand GPU cloud pricing as of December 2025.}
\small
\begin{tabular}{@{}p{2cm} p{3cm} p{2.5cm} p{2.5cm} p{6cm}}
\toprule
\textbf{Provider Category}  & \textbf{Provider Example} & \textbf{A100 (40GB/80GB) Price} & \textbf{H100 (80GB) Price} & \textbf{Specifics}\\
\midrule
\multirow{5}{4em}{Specialist GPU Cloud Platforms} & Thunder Compute & $\sim$\$0.78/hr & $\sim$\$1.99/hr	& Ultra-low pricing \\ 
\vspace{4em}
& Vast.ai & $\sim$\$0.67-\$1.47/hr & $\sim$\$1.61-\$1.87/hr & Marketplace, prices vary by host/availability \\ 

& RunPod & $\sim$\$1.39/hr & $\sim$\$1.99-\$2.69/hr & Community Cloud option is cheaper \\

& Lambda Labs & $\sim$\$1.79/hr & $\sim$\$2.99/hr & Enterprise grade hardware with simple pricing \\
\midrule
\multirow{5}{4em}{Hyperscalers} & Google Cloud (GCP) & $\sim$\$1.57/hr (spot) & $\sim$\$3.00/hr & Good on-demand rates, especially for spot \\
\vspace{4em}
& 	Amazon Web Services (AWS) & (Not listed) & $\sim$\$3.90/hr & Competitive after mid-2025 price cuts\\ 
& 	Microsoft Azure & \$3.67 & $\sim$\$6.98/hr & Higher on-demand rates \\
& 	Oracle Cloud & \$4 & $\sim$\$10.00/hr & Priced per 8-GPU bare metal node \\

\bottomrule
\end{tabular}
\label{tab:on_demand_gpu_pricing}
\end{table*}

For the Kinyarwanda ASR hackathon and similar short-term projects, using a specialist provider's on-demand or spot cloud GPU instances offers the best value. 
On the opposite, for medium-term projects (1-3 years) with more consistent workloads, i.e., more than 60\% continuous utilization, exploring reserved instances or commitments over corresponding time frames on any platform can yield significant discounts (30-60\%). 

\begin{table*}[!ht]
\centering
\caption{Domestic AI HPC vs. Cloud compute rental for AI model development.}
\small
\begin{tabular}{@{}p{2.5cm} p{7cm} p{7cm}}
\toprule
\textbf{Feature} & \textbf{Buying - 8xH100 System} & \textbf{Renting - 8xH100 Cloud Instance}\\
\midrule
Upfront Cost & \textbf{\$300K-\$500K+} (Hardware and initial setup.) & Potentially \textbf{\$0}-Minimal setup fees \\
\midrule
Operational Costs & Power, cooling, data center space, enterprise support and maintenance contracts, specialized DevOps and IT staff (up to \textbf{\$200K} annually) & Hourly/monthly fees, potential data egress costs\\
\midrule
Flexibility	& Low; locked into hardware for 3-5 years & High; scale up/down instantly; access to latest accelerator hardware and technologies\\
\midrule
TCO Break-even & Roughly \textbf{\$800K-\$1M}; requires over 60-70\% continuous utilization to be cheaper than renting over 3 years & Cost-effective for any utilization level below 60\% \\
\midrule
Sovereignty & High control; data stays in-house/local data center; local security standards apply & Depends on provider/region; major clouds offer data residency options\\
\bottomrule
\end{tabular}
\label{tab:app_cost_benefit_analysis}
\end{table*}

\subsubsection{Risks and Mitigation Strategies}
For AI development, choosing between buying and renting compute also involves managing distinct sets of risks that impact financial health and operational success. 
The primary risk of buying is rapid technology obsolescence (e.g., Ampere GPUs being replaced by Hopper and then Blackwell GPUs within 5 years), which locks significant capital into depreciating assets and imposes a heavy operational burden for maintenance and scaling. 
These risks are best mitigated by adopting a hybrid strategy, securing strong local maintenance contracts, and using hardware financing to preserve working capital.

Conversely, the main risks of renting revolve around cost volatility, ``bill shock'' from high data egress fees, and potential vendor lock-in or resource scarcity for high-demand advanced GPUs. 
These issues can be managed by implementing strong finance and DevOps, i.e. \textit{FinOps} practices, leveraging reserved instances for predictable workloads, utilizing multi-cloud strategies across specialist platforms and hyperscalers, and employing checkpointing software for interruptible spot instances to name a few.\\

\end{document}